\documentclass[a4paper,11pt]{article}

\usepackage{jheppub} 

\usepackage[T1]{fontenc} 

\usepackage{epsfig}
\usepackage{latexsym}
\usepackage{graphicx}
\usepackage{epstopdf}  
\usepackage[justification=RaggedRight]{caption}
\usepackage{amssymb}
\usepackage{subfig}
\usepackage{color}
\usepackage{multirow}
\usepackage{color}
\usepackage{rotating}
\usepackage{ifthen}
\usepackage{epsfig}
\usepackage{booktabs} 
\usepackage{hyperref} 
\usepackage{hyperref}
\hypersetup{
    colorlinks,%
    citecolor=blue,%
    filecolor=blue,%
    linkcolor=blue,%
    urlcolor=blue
}
\def\be{\begin{equation}}
\def\ee{\end{equation}}
\newcommand{\ds}{{\sffamily DarkSUSY}}

\def\eps{{\cal E}}

\newcommand{\bwt}{\begin{widetext}}
\newcommand{\ewt}{\end{widetext}}
\newcommand{\bdm}{\begin{displaymath}}
\newcommand{\edm}{\end{displaymath}}
\newcommand{\bea}{\begin{eqnarray}}
\newcommand{\eea}{\end{eqnarray}}

\newcommand{\vmin}{\ensuremath{v_\mathrm{min}}}

\newcommand{\bvobs}{\ensuremath{\mathbf{v}_\mathrm{obs}}}

\newcommand{\bu}{\ensuremath{\mathbf{u}}}  
\newcommand{\bv}{\ensuremath{\mathbf{v}}}  
\newcommand{\bV}{\ensuremath{\mathbf{V}}}  

\newcommand{\eone}{\ensuremath{\hat{\mathbf{\varepsilon}}_1}}  
\newcommand{\etwo}{\ensuremath{\hat{\mathbf{\varepsilon}}_2}}  
\newcommand{\beqra}{\begin{eqnarray}}
\newcommand{\eeqra}{\end{eqnarray}}
\newcommand{\beq}{\begin{equation}}
\newcommand{\eeq}{\end{equation}}

\newcommand{\supb}{{\sffamily SuperBayes }}

\title{Dark matter signals at the LHC: forecasts from ton-scale direct detection experiments}

\author[a]{Giorgio Arcadi,}
\author[a]{Riccardo Catena}
\author[b]{and Piero Ullio}

\affiliation[a]{Institut f\"ur Theoretische Physik, Friedrich-Hund-Platz 1, 37077 G\"ottingen, Germany}
\affiliation[b]{SISSA and INFN, Sezione di Trieste, Via Bonomea, 265, 34136 Trieste, Italy}

\emailAdd{giorgio.arcadi@theorie.physik.uni-goettingen.de}
\emailAdd{riccardo.catena@theorie.physik.uni-goettingen.de}
\emailAdd{ullio@sissa.it}

\abstract{The complementarity between dark matter searches at colliders and in underground laboratories is an extraordinarily powerful tool in the quest for dark matter. In the vast majority of the analyses conducted so far these dark matter detection strategies have been profitably combined either to perform global fits in the context of certain particle physics models (e.g. the CMSSM) or to estimate the prospects for a direct dark matter detection given the LHC potential of discovering new physics beyond the Standard Model.
In this paper we propose an alternative strategy to combine direct and collider dark matter searches: employing the potential of the upcoming generation of 1-ton direct detection experiments, we show that for certain supersymmetric configurations it is possible to translate the information encoded in an hypothetically discovered direct detection signal into classes of expected signals at the LHC. As an illustrative application of our method, we show that for a 60 GeV neutralino thermally produced via resonant annihilations and identified by a 1-ton direct detection experiment, our approach allows to forecast a clearly identifiable prediction for a LHC final state involving three leptons and missing energy. The strategy presented in this paper to systematically translate a direct detection signal into a prediction for the LHC has the potential to significantly strengthen the complementarity between these two dark matter detection strategies.}

\begin{document} 
\maketitle
\flushbottom

\section{Introduction}

The nature of the dark matter (DM) component of the Universe remains unknown. One of the most popular solutions to this puzzle is the scenario in which DM is made of a beyond-Standard-Model (SM) particle which is stable, in thermal equilibrium in the early Universe and non-relativistic at chemical decoupling; the relic density for such a state scales approximately with the inverse of its  pair annihilation rate into lighter SM particles and matches the abundance of DM as determined from cosmological measurements for a weak-interaction-strength coupling: this is the essence of the  celebrated paradigm for weakly  interacting massive particles (WIMPs) as DM candidates. On top of the fact that there are numerous extensions of the SM naturally embedding  a WIMP DM state, to some extent the popularity of this framework relies on its very rich phenomenology, with a set of complementary detection strategies stemming from the framework itself on rather model-independent bases: One in general expects a tight correlation between the annihilation rate of DM in the early Universe (at a temperature of about one twentieth of the particle mass) and that in dark halos today (at relative velocities as small as 10 to 1000 km~s$^{-1}$) giving rise to a non-standard astrophysical source of photons, leptons or antimatter through which indirectly detect the DM particles. On the other hand, applying CP symmetry arguments to the annihilation process, one expects to produce DM states in inelastic scatterings of SM particles at high energy colliders, such as the LHC. Finally, based on crossing symmetry, one generally predicts that the weak coupling to ordinary matter in the early Universe corresponds to a small but still significant elastic scattering cross section for DM particles on ordinary matter, through which directly detect the DM WIMPs forming the halo of the Milky Way. Direct, indirect and collider searches for WIMPs have then been primary goals in the experimental  effort  towards the identification of the nature of DM in the latest twenty years or so, having reached today the stage when, in all three techniques, experiments have the capability to single out or exclude viable DM models.

In this paper we investigate some aspects of the complementarity between direct detection (DD) and collider searches. The goal is to investigate the potential of the upcoming generation of DD experiments, taking as reference a setup resembling  the 1-ton realization of the XENON detector, and cross correlate the information encoded in a hypothetically discovered signal with the expected signals at the LHC. Modeling the coupling of DM and SM states at an effective operator level, such as in the recent analysis of, e.g., ~\cite{Fox:2011pm} one can drive a direct link between the two detection strategies; on the other hand, when considering explicit WIMP models cross symmetry arguments rarely apply, with the couplings ensuring the appropriate relic density most often not corresponding to those setting the WIMP-nucleus elastic scattering cross sections and those responsible for the production of DM states in LHC's pp collisions. We choose then here to focus on a definite scenario, assuming that the DM particle is the lightest neutralino in the minimal supersymmetric extension to the SM (MSSM). This choice is motivated by the fact that, with a number of degrees of freedom which is more than twice those of the SM and the large number of free parameters involved (we will refer to a simplified version of the so-called phenomenological MSSM, or pMSSM, allowing for 12 free parameters), this framework is general enough to embed several different mechanisms to provide the correct relic abundance and/or a scattering rate at a detectable level. At the same time, the neutralino is the most closely investigated WIMP DM candidate within the most widely studied framework for searches for physics beyond the SM at colliders (as well as in rare flavor processes): we will then exploit the several well-known patterns for this DM state as criteria to explore the complex parameter space and interpret results, as well refer to a set of well-established search channels to extract guidelines which are actually more general than for the MSSM itself.

Our analysis starts with the choice within the MSSM of a set of few benchmark models
representative of WIMP DM candidates which, in the plane spin-independent (SI) elastic scattering cross section per nucleon versus DM mass, appear below the region excluded by DD searches -- with the most relevant limit stemming from the null result recently reported by the XENON100 experiment~\cite{xenon:2012nq} -- but still giving a sizable number of scattering events for a 1-ton detector. All benchmarks are selected checking that the lightest Higgs boson in each model matches the Higgs-like state recently discovered at the LHC, and fulfilling current constraints on beyond-SM states from rare process like $b \rightarrow s \gamma$ and $B_s \rightarrow \mu^+ \mu^-$ as well as general trends from direct SUSY searches at colliders.
For each benchmark, first a set of mock data for the recoil spectrum in a 1-ton detector is simulated; such spectrum depends, on top of the SI scattering cross section and the DM mass, on the phase-space distribution of DM particles. Uncertainties on the latter have been recently investigated in Refs.~\cite{Catena:2009mf,Catena:2011kv}, implementing a large set of dynamical observables for the Milky Way and under the simplifying (but reasonable) approximations of spherical symmetry and isotropy; we will exploit results of such analysis in our approach here (complementary methods to study the local DM distribution have been discussed for instance in Refs.~\cite{Iocco:2011jz,Garbari:2011dh,Pato:2010yq,Salucci:2010qr,deBoer:2010eh,Hansen:2005yj}). Our simulated DD data samples consist in a number of measured recoil energies varying from approximately 30 to 190, depending from the benchmark model under analysis. Before analyzing these data, a statistical test has been performed on the simulated data samples to assess whether they provide a faithful realization of their true underlying distribution. This aspect has been addressed with great care since statistical fluctuations in the distribution of the observed recoil energies can potentially affect the DM mass and scattering cross section determinations significantly, in particular for heavy DM candidates with low scattering cross sections~\cite{Strege:2012kv}.
Mock data are then analyzed employing a Bayesian approach supported by a Markov Chain Monte Carlo (MCMC) Likelihood scanning technique of the parameter space defined by both the particle physics model parameters and the parameters related to the DM phase-space distribution function. The results of this analysis regarding the preferred values for the DM mass and scattering cross section will be presented in terms of Bayesian credible regions. 
Having derived through this procedure the accuracy within which the relevant DM properties, namely mass and scattering cross section, are reconstructed and their mapping on the MSSM parameter space, we take this input to outline the most suitable collider search strategies and argue next future detection prospects. This second part of our analysis has been performed by focusing on the regions of the DM parameter space characterized by the highest statistical weight (i.e. with the largest Bayesian posterior probability density resulting from our DD analysis) and simulating for a subset of models, sampled within these regions, the ATLAS detector response in its configuration with 7 TeV center of mass energy. This allowed us, in particular in the case of one of the benchmark investigated here, to effectively translate the simulated DD signals studied in the first part of this work into definite predictions for certain LHC observables. This procedure defines a method to efficiently exploit the upcoming generations of 1-ton DD detectors to forecast valuable predictions for the LHC and it constitutes the main achievement of this paper.
The paper is organized as follows: in section \ref{sec:pp} we introduce the particle physics framework adopted in our analysis; in section \ref{sec:bm} we provide a description of the benchmark points studied in this paper and review the criteria which brought to their selection. Section \ref{sec:dd} is instead devoted to a study of the capability of a ton-scale DD experiment in reconstructing mass and SI scattering cross section of the DM candidates associated with these benchmark points. In section \ref{sec:lhc} we will translate these informations on the DM mass and scattering cross section into a prediction for certain classes of LHC final states while our conclusions will be summarized in section \ref{sec:conclusions}.  

\section{Particle physics framework}
\label{sec:pp}
The aim of this paper is to develop a method to translate an hypothetically discovered DD signal into LHC observable quantities ({\it e.g.} missing energy distributions). Obviously, since the DD technique is directly sensitive only to the DM mass and scattering cross section, while LHC observables generically depend from different (combinations of) couplings and masses, assumptions regarding the underlaying particle physics model have to be made in order to achieve this goal. Tackling this task relying on an effective field theory of the DM interactions would lead to a likely oversimplified picture of the actual possible correlations among DD scattering cross sections and LHC missing energy distributions. We therefore focus on a more general (but still manageable) particle physics framework, namely the MSSM and, more precisely, on its phenomenological realization \cite{Djouadi:1998di} (pMSSM), which for completeness we briefly review in the following.

\subsection{pMSSM}
\label{sec:pMSSM_setup}

The most general realization of the pMSSM features 22 parameters defined at the Electro-Weak (EW) scale, namely:
\begin{itemize}
\item $M_1$, $M_2$ and $M_3$ parameters corresponding to the gaugino masses;
\item $\mu$, $m_A$ and $\tan\beta$ parameters describing the Higgs/Higgsino sector;
\item sfermion soft-SUSY breaking mass parameters $m_{\tilde{q}}$, $m_{\tilde{u}_R}$, $m_{\tilde{d}_R}$, $m_{{\tilde{l}}_L}$, $m_{\tilde{e}_R}$,  equal for first and second generation;
\item  trilinear soft terms $A_u$, $A_d$ and $A_e$ equal for the first and second  generation;
\item third generation soft mass parameters $m_{{\tilde{q}}_{3L}}$, $m_{\tilde{t}_R}$, $m_{\tilde{b}_R}$, $m_{{\tilde{l}}_{3L}}$, $m_{\tilde{\tau}_R}$;
\item third generation trilinear soft terms $A_t$, $A_b$ and $A_\tau$;
\end{itemize}
with the further assumption that no additional sources, with respect to the SM,  of flavor and CP violation are present. The exploration of such large parameter space is, however, computationally very demanding; for this reason we have considered a few limits reducing its dimensionality, paying attention however to still cover all the scenarios that are, possibly, phenomenologically relevant.

In our implementation of the pMSSM we have neglected contributions to all the observables from squarks of the first two families fixing their mass at about a few TeV, therefore decoupling them from the rest of the spectrum. In addition, their corresponding trilinear $A-$terms have been set to zero. This choice is motivated by the fact that current null results of SUSY searches at LHC constraint more severely the masses of the squarks of the first two generations (compared to the third generation masses), ruling out in a few scenarios values below approximately 1 TeV \cite{Aad:2011ib,Chatrchyan:2011zy}. In addition to these simplifications, we have further reduced the parameter space in the sfermion sector by assuming $m_{\tilde{l}}=m_{{\tilde{l}_L}}=m_{\tilde{e}_R}$, $m_{\tilde{l}_{3}}=m_{{\tilde{l}}_{3L}}=m_{\tilde{\tau}_R}$, and $m_{{\tilde{q}}_{3R}}=m_{\tilde{b}_R}=m_{\tilde{t}_R}$. Finally, we have also set  $A_b=A_\tau$. This reduces the pMSSM parameter space to a set of 12 free parameters.

\subsection{Constraints on the pMSSM}
The pMSSM benchmark points for which our method will be tested in the next sections have been constructed requiring that they pass a variety of complementary constraints. We have first of all required that the generated spectra provide a CP-even Higgs boson mass in agreement with the recent LHC discovery \cite{atlas:2012gk} of a neutral boson of mass of around 125.5 GeV and couplings compatible with the ones of the SM Higgs boson (there are slight indications, not statistically decisive, of some deviations from the standard values, see for details \cite{Giardino:2012dp, Carmi:2012in}). The MSSM setup most likely in agreement with this result is the so-called decoupling regime \cite{Haber:1995be} in which the light CP-even Higgs state $h$ has SM-like couplings and is required to have mass in agreement with the experimental determination, while all the other Higgs states are degenerate in mass and lie at a sensitively higher scale. In this regime the Higgs mass $m_h$ is approximatively given by the sum of tree-level term and stop-top loop contributions:
\begin{equation}
\label{eq:1loophiggs}
m_h \approx m_{\rm Z} \cos 2 \beta + \frac{3 G_F}{\sqrt{2} \pi^2} m_t^4 \left(\log\left(\frac{M_{s}^2}{m_t^2}\right)+\frac{X_t^2}{M_s^2}\left(1-\frac{X_t^2}{12 M_s^2}\right)\right)
\end{equation}
where $X_t \equiv A_t -\mu \cot \beta$, namely the stop mixing parameter, and $M_s =\sqrt{m_{\tilde{t}_1} m_{\tilde{t}_2}}$ with $\tilde{t}_1$ and $\tilde{t}_2$ being the two stop mass eigenstates. 
Matching the experimental value of the Higgs mass is rather challenging for the MSSM. 
Indeed it requires a large contribution from loop-corrections which achieve the correct value only in the case the stop mass scale lies above around 3 TeV unless the stop mixing parameter $X_t = A_t -\mu \tan\beta$ fulfills the so called maximal mixing condition~\cite{Baer:2011ab,Arbey:2011ab}, namely $X_t \approx \sqrt{6} M_s$ \footnote{Milder constraints are obtained in the minimal extension of the MSSM, the so called NMSSM, in which the Higgs couples with and additional superfield which provides an additional one-loop contribution to its mass \cite{Hall:2011aa}.}. In this work (in particular in the DD analysis of section \ref{sec:dd}) we have imposed the requirement $123 \le m_h \le 129$ GeV, taking into account the experimental uncertainties as well as the ones affecting the numerical computation.

We just mention that correct values of the Higgs mass can be also achieved by configurations apart from the decoupling regime, featuring light masses, {\it i.e.} close in mass to the LHC candidate, for the other Higgs states  (see e.g. \cite{Arbey:2012dq} for a review). However these are severely constrained by the searches of CP-odd Higgs decays into tau pairs~\cite{Aad:2011rv,Chatrchyan:2012vp} and, as we will clarify in the following, by the limits from $Br(B_s \rightarrow \mu^+ \mu^-)$. We finally notice that light DM scenarios may influence Higgs phenomenology allowing for the decay $h \rightarrow \chi \chi$. 

The invisible decay width of the Higgs has been tested by some early studies based on multijet + missing $E_T$ events originated by the associated production of the Higgs with a gauge boson or jet. This kind of studies put an upper bound on the invisible branching ratio of the Higgs of the order of 0.5 \cite{Fox:2011pm}. More recent studies are aiming at infering the viability of a non zero $Br(h \rightarrow \chi \chi)$ by fitting the current data relying on Higgs searches. A study of this type can be found e.g. in Ref.~\cite{Giardino:2012ww} in which a fit of the current data has been performed by adding to the decay channels into visible states a branching ratio into two DM particles made free to vary. The outcome of such analysis is that $BR(h \rightarrow \chi \chi) > 0.4$ is excluded at the 95$\%$ confidence level with the best fit pointing towards values close to zero.

The second set of bounds considered in the benchmark points selection is related to the searches for FCNC processes. We focused on the three most important observables, namely the anomalous magnetic moment $a_{\mu}$, the decay processes $b \rightarrow s \gamma$ and $B_s \rightarrow \mu^{+} \mu^{-}$. Regarding the anomalous moment of the muon, the SUSY contributions come from neutralino-slepton loops and chargino-sneutrino loops \cite{Moroi:1995yh} with the latter typically dominating \cite{Calibbi:2011dn}. 

Contributions to the $b \rightarrow s \gamma$ processes are instead induced by charged Higgs-top loops and chargino-stop loops \cite{Lunghi:2006hc} which have to be combined with the SM contribution, $Br_{\rm SM}(b \rightarrow s \gamma) \simeq 3.0 \times 10^{-4}$. While the charged Higgs-top loops always add constructively to the SM contribution, the behavior of chargino/stop loops depends on the sign of the quantity $Re(\mu A_t)$. Having fixed the $\mu$ parameter to a positive value,  they add destructively to the SM one when $A_t < 0$ and constructively when $A_t >0$ . We also remark that the chargino-stop loops contribution is enhanced at high $\tan\beta$. 

Finally, regarding $B_s \rightarrow \mu^{+} \mu^{-}$, the additional contributions with respect to the SM arise from neutral heavy Higgs loops and are proportional to $\tan^3\beta$ and to $A_t \mu$. This dependence on $\tan^3\beta$ and the heavy Higgs mass makes, as already stated, this process also relevant for the Higgs phenomenology. In our analysis we made use of the numerical package SuperIso \cite{Mahmoudi:2007vz,Mahmoudi:2008tp} for computing the rates of these three process imposing the range $[2.77,4.37] \times 10^{-4}$ for the values of $Br(b \rightarrow s \gamma)$ and upper bounds of $4.79 \times 10^{-9}$ for $\delta a_\mu$ and $4.5 \times 10^{-9}$ for $Br(B_s \rightarrow \mu^+ \mu^-)$~\cite{Chatrchyan:2012rg, Aaij:2012ac}. 

\section{Benchmark points selection}
\label{sec:bm}
Having specified the particle physics framework and the experimental constraints imposed on the associated parameter space, we can now concentrate on the selection of the benchmark points for which our method to combine DD and collider searches will be tested.

\subsection{Guiding principles}
The guiding principles driving our benchmark points selection are: the requirement of a correct DM relic density and the necessity of a potentially observable SI scattering cross section. To clarify the reasons behind the choice of the specific benchmark points introduced in the next section, we therefore find instructive to first briefly review a few relevant aspects of the DM thermal production mechanism and of the calculation leading to the explicit form of the SI scattering cross section. 

In the context of the pMSSM there are rather different regimes in which the thermal production mechanism is successfully realized, providing a value for the neutralino relic density compatible with CMB observations. The details of the DM thermal production depend from whether the lightest neutralino is higgsino- and wino-like, or bino-like. In the former case, since higgsino and winos have unsuppressed couplings to W bosons via their charged counterparts, the pair annihilation rate is very large and the relic density matches the cosmological DM density only going to the heavy mass regime, at about 1.4 and 2.7 TeV for, respectively, a pure higgsino and a pure wino \cite{Hryczuk:2010zi}.  Such a regime is however not favorable for a DD analysis and will be therefore discarded here in the benchmark points selection process. Moving to bino-like DM, we have an opposite situation, since bino-like DM couples to fermions only but the s-wave pair annihilation rate is helicity suppressed, namely proportional to the square of the mass of the final state fermion. As a consequence the value of the DM relic density exceeds the experimental limit with the exception of some peculiar regions of the parameter space, where definite relations among pMSSM parameters exist. Some of these configurations are also present in the CMSSM, thus for simplicity we will refer to them in the jargon borrowed from this class of models. Bino-like DM candidates can be cosmologically viable in the following cases:
\begin{itemize}
\item In the so-called ``bulk region'', where DM pair annihilations are mediated by a sfermon with mass of the order of 100-200 GeV. These SUSY configurations are already ruled out in the CMSSM by current LHC limits, but still partially allowed in the pMSSM if the light sfermion is a slepton.
\item In the so called ``Higgs funnel region'', which is characterized by an enhancement of the DM pair annihilation rate due to the presence of an s-channel resonance. In the CMSSM this can happen via resonant annihilations mediated by a CP-odd Higgs boson, when $m_A \simeq 2 m_\chi$. This configuration can be also realized in the pMSSM. This occurs for small DM masses through resonances associated to the Z boson and to the light CP-even Higgs. The latter configuration is particularly interesting in light of the recent discovery of a Higgs-like state at LHC, being in this case DM and Higgs physics deeply connected.
\item In the case in which the DM is a mixture of a bino and a small higgsino or wino component. The first case also occurs in the CMSSM in the so-called ``focus point region'' while the pMSSM admits both configurations, which in this context are also dubbed as ``well-tempered neutralinos" \cite{ArkaniHamed:2006mb}.
\item  In the so-called ``coannihilation scenario'', namely when one or more particles almost degenerate in mass with the DM contribute to keep the DM in chemical equilibrium in the early Universe. Remarkably, coannihilating particles evade most of the current SUSY searches because the low mass splittings disfavor the detection of their decay products, after an eventual production at collider.
\end{itemize}

\begin{figure}[t]
\begin{center}
\includegraphics[width=9 cm, height= 7 cm, angle=360]{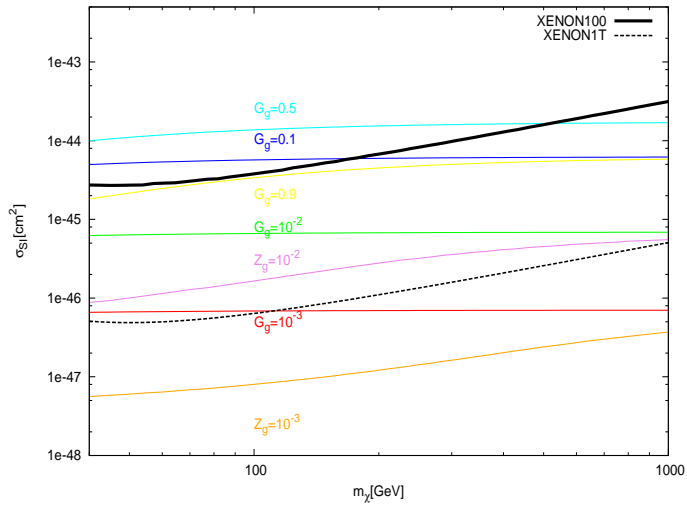}
\end{center}
\caption{Curves at the same values of the DM bino (higgsino) fraction $G_g$ ($Z_g$) in the plane $(M_\chi, \sigma_{SI})$. These curves have been obtained by considering a simplified model featuring only $M_1$ and $\mu$ as free parameters while all the remaining masses are fixed at about few TeV. In addition $m_h$ and $\tan\beta$ have been fixed to, respectively, 126 GeV and 10. The solid and the dashed black line are, respectively, the XENON100 limit and the XENON1T projected sensitivity. }
\label{fig:piero1}
\end{figure}

Concerning the SI scattering cross section, its general form in the small momentum transfer limit can be found for instance in Ref.~\cite{Ellis:2005mb}\footnote{The expression there reported may not be strictly valid in case of coannihilations and need to be eventually corrected according to~\cite{Drees:1992rr,Drees:1993bu}.}. In this expression one can identify two main contributions. A first contribution, depending on the SUSY parameters involved in the Higgs sector, comes from the t-channel interaction mediated by a CP-even Higgs. In the decoupling regime only the light Higgs state $h$ is relevant with the coupling scaling with the gaugino-higgsino mixing, being zero for a pure bino or a pure higgsino. This effect can be seen in Fig.~\ref{fig:piero1}, where, in a simplified model featuring only $M_1$ and $\mu$ as free parameters, we show the isocontour for the bino (higgsino) fraction $G_g$ ($Z_g$) in the plane DM mass versus SI scattering cross section.
The second contribution comes from interactions mediated by squarks. This is therefore suppressed, compared to the former, by the squark mass scale. There is however a possible exception occurring for very low mass splittings between a squark and the DM particle.  A further enhancement of the contribution to the SI scattering cross section coming from squark mediated interactions is obtained assuming a sizable mixing between left and right-handed states. In our particle physics framework this condition can be fulfilled by third family squarks \footnote{This further motivates our choice of limiting to third generation squarks the possibility of being at low mass}. 

Motivated by these considerations we decided to choose one benchmark point in the ``Higgs funnel region'' (where DM is relatively light and linked to the Higgs physics) and three points representative of different ``coannihilation scenarios'', evading these configurations most of the current LHC limits. Moreover, to equally explore the two regimes of the SI scattering cross section, we selected two benchmark points with DM-quark coupling dominated by Higgs mediated interactions and two points associated instead with interactions dominated by a squark exchange. The properties of these four benchmark points will be fully illustrated in the next section. 

\subsection{Four benchmark points}
\begin{figure}[t]
\begin{center}
\includegraphics[width=9 cm, height= 8cm]{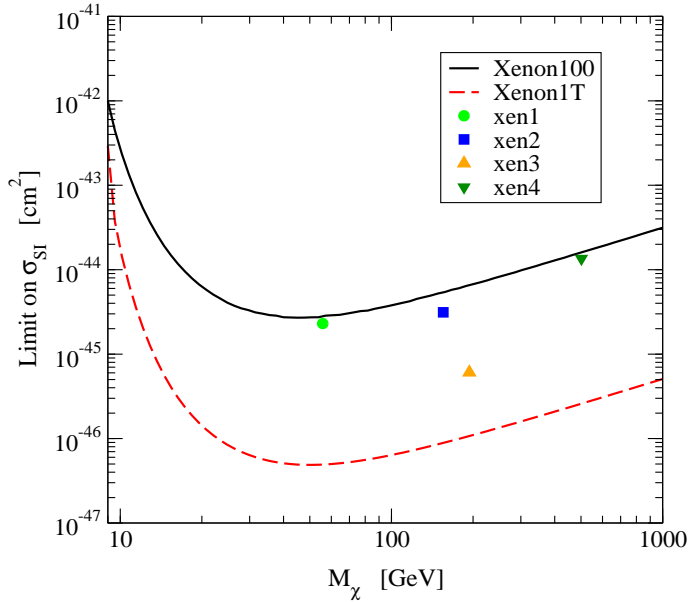}
\end{center}
\caption{The four benchmark points studied in this paper are shown together with the present Xenon exclusion limit and an estimation for the expected sensitivity of the Xenon experiment in its 1-ton configuration.}
\label{fig:bmall}
\end{figure}
The first benchmark, labeled by {\bf{xen1}}, has been selected within  the Higgs funnel region and is characterized by a SI scattering cross section dominated by Higgs mediated interactions, being the squark masses very large compared to the DM mass at this benchmark point. For {\bf{xen1}} both the thermal production and the DD signal are controlled by the neutralino-neutralino-Higgs vertex, which is in turn mostly sensitive to the pMSSM parameters $M_1$, $M_2$ and $\mu$, from which the neutralino composition depends. {\bf{xen1}} features a DM mass of approximately 56 GeV, therefore within the mass range for which DD experiments have the highest sensitivity. 

The second benchmark point, {\bf xen2}, generates a DM candidate with mass close to 150 GeV whose relic density matches the observed one through efficient DM-stau coannihilations. SI interactions mediated by squarks are suppressed in this scenario since, analogously to {\bf{xen1}}, squarks are very heavy at this benchmark point. We selected this point with the aim of investigating collider signals possibly induced by light sleptons. In the case of  {\bf xen2} the relic density is very sensitive to the stau mass while, as for the case of  {\bf{xen1}}, the SI scattering cross section is mostly controlled by the parameters $M_1$, $M_2$ and $\mu$.

Finally, we have considered two benchmark points, namely {\bf xen3} and {\bf xen4}, corresponding to scenarios where DM is thermally produced through coannihilations with light squarks of the third family. The first of them, namely {\bf xen3}, has been designed to have a very low value of the SI cross section and a mass close to the current exclusion limits for this scenario ({\it i.e.} 200 GeV), including the bounds from direct SUSY searches at LHC. The benchmark point {\bf xen4} is instead the heaviest benchmark considered in this paper, having a mass close to 500 GeV. For the benchmark points {\bf xen3} and {\bf xen4} the most relevant parameters in the calculation of the relic density and of the SI scattering cross section are $M_1$, stop and sbottom masses, the parameters determining the mixing between left and right-handed squark states (see our discussion in the previous section) and, only in the case of {\bf xen4}, the $\mu$ parameter, which in this case is not significantly larger than $M_1$ therefore producing a small higgsino fraction in the lightest neutralino composition. 

The main features of the four benchmark points are summarized in Tab.~\ref{tab:spectra2} while Fig.~\ref{fig:bmall} shows the four benchmarks in the plane DM mass versus SI cross section together with the present Xenon exclusion limit and an estimation of the expected sensitivity of the Xenon experiment in its 1-ton configuration. Contrary to the case of an effective field theory approach to the DM interactions where for every given higher dimensional operator there are just two relevant parameters, {\it i.e.} the DM mass and the scale of new physics which suppresses the operator considered, the four benchmark points studied here have properties depending from a slightly larger number of parameters. This allows to explore less drastically simplified scenarios, keeping at same time under control the degree of complexity of the analysis.

\subsection{Overview of our approach}

Having defined the benchmark points studied in this paper, we can now introduce the method which we propose (and apply in the next sections) to translate an hypothetically discovered DD signal into a prediction for certain LHC final states. The basic idea is that, although the DD technique is directly sensitive only to the DM mass and scattering cross section, indirectly this class of experiments has also the potential to constraint the parameters, or certain combinations of parameters, which most crucially enter in the calculation of the DM scattering cross section. These parameters are $M_1$, $M_2$ and $\mu$ in the case of {\bf xen1}, for instance. Then one can focus on those LHC final states whose associated missing energy distribution (possibly related to DM production) depends mostly from the very same parameters. In the case of {\bf xen1} one of these processes would be the direct production of a $\tilde{\chi}^0_2\tilde{\chi}^{\pm}_1$ pair leading to a final state with three leptons plus missing energy. So, evaluating the missing energy distribution associated with these final states for the model parameters ``favored'' by the outcome of a simulated 1-ton DD experiment, we can work out a genuine DD driven prediction for the LHC. In the next sections will apply this idea to {\bf xen1}, {\bf xen2}, {\bf xen3} and {\bf xen4}, starting from the Bayesian simulation of a ton-scale DD experiment. 

\begin{table}[t]
\begin{center}
\begin{tabular}{|c|c|c|c|c|c|c|c|c|c|}
\hline
benchmark & $m_{\chi}$ & $m_A$ & $ \mu $ & $m_{\tilde{t_1}}$ & $m_{\tilde{b_1}}$ & $m_{\tilde{g}}$ & $m_{\tilde{\tau}}$ & $\sigma_{SI}$ & N\\
\hline
xen1 & 56 & 890 & 291 & 557 & 735 & 815 & 165 & $2.3 \times {10}^{-45}{\mbox{cm}}^2$ & 170 \\
\hline
xen2 & 155 & 1550 & 353 & 1715 & 1777 & 2254 & 163 & $3.12 \times {10}^{-45}{\mbox{cm}}^2$ & 129 \\
\hline
xen3 & 194 & 2900 & 1500 & 340 & 209 & 2278 & 732 & $6.1 \times {10}^{-46}{\mbox{cm}}^2$ & 24 \\
\hline
xen4 & 477 & 2780 & 718 & 527 & 673 & 665 & 656 & $1.35 \times {10}^{-44}{\mbox{cm}}^2$ &187 \\
\hline
\end{tabular}
\end{center}
\caption{Summary table of the benchmark points. In the last column we report the number of recoil energies simulated in the DD analysis of section \ref{sec:dd}.}
\label{tab:spectra2}
\end{table}%

\section{Direct detection}
\label{sec:dd}

In this section we will investigate the capability of a ton-scale DD experiment in reconstructing mass and SI scattering cross section associated with the pMSSM benchmarks previously introduced. DD experiments seek to measure the nuclear recoil energy deposited in an underground detector when an incident $\chi$ particle of the Milky Way DM halo interacts with a nucleus of the target material. In the case of SI $\chi$-nucleus interactions the differential cross section for this process can be written as
\begin{equation}
\frac{d\sigma}{dQ} = \sigma_{SI} \frac{M_N}{2\,M_n^2\,|\bu|^2} A^2 {\mathcal F}_N^2(Q)\,,
\label{dsigma}
\end{equation}
where $Q$ is the energy transferred to the nucleus of mass $M_N$ from a $\chi$ particle moving with velocity $\bu$ in the detector rest frame and $A$ is the nucleus mass number. The form factor ${\mathcal F}_N(Q)$ accounts for the internal structure of the nucleus while $\sigma_{SI}$ is the $\chi$-nucleon SI scattering cross section at zero momentum transfer and $M_n = M_\chi m_p/(M_\chi + m_p)$ the $\chi$-nucleon reduced mass. The differential rate of scattering events per unit detector mass in a DD experiment is given by the convolution of the incident flux of $\chi$ particles with the differential cross section (\ref{dsigma}):
\begin{equation}
  \frac{dR}{dQ} = \frac{\sigma_{SI}}{2\,M_{\chi}M_n^2} A^2 {\mathcal F}_N^2(Q)
  \int_{|\bu| > \vmin} d^3u\; \frac{\rho_\chi  \cdot f_\chi(\bu,t)}{|\bu|} \,,
\label{eq:ddrate2}
\end{equation}
where the lower limit in the integral is the minimum velocity required for a $\chi$ particle to deposit the energy $Q$. In the case of an elastic scattering, such as for WIMP-nucleus interactions, the minimum velocity is given by
\beq
\vmin = \sqrt{\frac{Q M_N}{2 M_r^2}}\,.
\label{vmin-el} 
\eeq
In Eq.~(\ref{vmin-el}) $M_r = M_\chi M_N/(M_\chi + M_N)$ is the $\chi$-nucleus reduced mass while $M_{\chi}$ represents the mass of the $\chi$ particle. 

The product $\rho_\chi \cdot f_\chi\equiv F_h$ of the local DM density $\rho_\chi$ and velocity distribution $f_\chi$ is equal to the phase-space density of the halo DM particles $F_h$ evaluated at the detector and in its reference frame. The local $\chi$ phase-space density is computed assuming that the $\chi$ particles 
account for the entire DM component and applying Eddington's inversion formula~\cite{BT:1998} to the mass model for the Milky Way introduced in section \ref{gmod}. This model is characterized by a spherically symmetric and isotropic distribution of DM particles and it furthermore assumes spherical symmetry for the underlying Galactic gravitational potential $\Phi(r)$. In this limit, Eddington's formula~\cite{BT:1998} gives an one-to-one correspondence between the DM halo density profile $\rho_h(r)$ and its phase-space density $F_h$:
\begin{equation}
   F_h(\eps) = \frac{1}{\sqrt{8} \pi^2} \left[ \int_0^{\eps} \frac{d^2\rho_h}{d\Psi^2} \frac{d\Psi}{\sqrt{\eps-\Psi}} + 
   \frac{1}{\sqrt{\eps}} \left(\frac{d\rho_h}{d\Psi}\right)_{\Psi=0} \right]\,,
\label{eq:edd}
\end{equation}
where we introduced the relative potential $\Psi(r) = - \Phi(r) + \Phi (r=R_{vir})$ and the virial radius $R_{vir}$, corresponding to the radius at which the DM halo is truncated. Under the assumptions considered here, $F_h$ is a function of the relative energy $\eps = - E + \Phi (r=R_{vir}) = - E_{\rm{kin}} + \Psi(r)$ only, with $E$ and $E_{\rm{kin}}$, respectively, the total and kinetic energies. The relative energy depends on the DM particle velocity and position in the Galaxy and therefore determines the space-time dependence of $F_h$ itself. To numerically compute $F_h$ at the local Galactocentric distance it is actually simpler to implement Eq.~(\ref{eq:edd}) by changing the integration variable from $\Psi$ to the radius of the spherical system $r$.  Then, the radial dependence of $\rho_h$ and $\Phi$ from the local position in the Galaxy out to $R_{vir}$ is the only information needed to calculate $F_{h}$. Further details on the implementation and validity of this approach can be found in \cite{Catena:2011kv}. In summary, Eddington's inversion formula allows us to express the integrand of Eq.~(\ref{eq:ddrate2}) directly as 
\begin{equation} 
  \rho_\chi  \cdot f_\chi(\bu,t) = F_h\left[\eps_0(\bv)\right] \,,
  \label{eq:rho-f}
\end{equation}
with $\bv(t)=\bvobs(t) + \bu$, being $\bvobs$ the velocity of the detector in the Galactic frame, where $F_h$ is computed, and $\eps_0$ the relative energy at the local Galactocentric distance. For $\bvobs$ we take the expression 
\begin{equation}
  \bvobs(t)= \bv_{\mathrm{LSR}} + \bv_{\odot,\mathrm{pec}} + \bV_\oplus(t) \,,
  \label{eq:coo1}
\end{equation}
where the Local Standard of Rest velocity $\bv_{\mathrm{LSR}}$ (namely the rotation of the Sun with velocity $\Theta_0$ around the Galactic center; calculable given a mass model for the Galaxy), the Sun's peculiar velocity $\bv_{\odot,\mathrm{pec}}$ and the motion of the Earth around the Sun $\bV_\oplus$ are given by
\beqra
&&\bv_{\mathrm{LSR}} = (0,\Theta_0,0) \,,\nonumber\\
&&\bv_{\odot,\mathrm{pec}} = (U_\odot,V_\odot,W_\odot) \simeq (10,5.2,7.2)\; {\rm km/s} \nonumber\\ 
&&\bV_\oplus(t) =  V_\oplus \left[\eone \cos{\omega(t-t_1)} + \etwo \sin{\omega(t-t_1)}\right] \,,
\label{eq:coo2}
\eeqra
with $V_\oplus=29.8$ km/s and $t_1=0.218$ is the fraction of the year before the Spring equinox. Finally, $\eone$ and $\etwo$ are the directions of the Earth's velocity at times $t_1$ and $t_1+0.25$ years. In Eq.~(\ref{eq:coo2}) all velocities are expressed in Galactic coordinates, where $\hat{\mathbf{x}}$ is the direction to the Galactic Center, $\hat{\mathbf{y}}$ the direction of disk rotation and $\hat{\mathbf{z}}$ identifies the North Galactic Pole.

\subsection{Galactic model}
\label{gmod}
The Galactic mass model adopted in the present analysis to compute $F_h$ has been extensively investigated in Refs.~\cite{Catena:2009mf,Catena:2011kv}. For completeness, we briefly summarize it in what follows. The model consists in two luminous mass components, namely the stellar disk and the Galactic bulge, and in a DM halo. The Galactic model parameters introduced in this section will be then treated as nuisance parameters in the Bayesian analysis of our pMSSM benchmarks performed in the next subsections. 

Regarding the stellar disk, we assume a mass density profile which in cylindrical coordinates $(R,z)$ with origin in the Galactic center is given by
\beq
\rho_d(R,z) = \frac{\Sigma_{d}}{2 z_{d}} \, e^{-\frac{R}{R_d}} \, \textrm{sech}^2\left( \frac{z}{z_d}\right)
\;\;\;\; {\rm{with}} \;\;\;\; R<R_{dm}\;,
\label{disk}
\eeq 
where $\Sigma_{d}$ is the central disk surface density, $R_d$ and $z_d$ are length scales in the radial and vertical directions, while $R_{dm}$ is the truncation radius of the disk. $R_{dm}$ is assumed to scale with the local Galactocentric distance $R_0$ according to the prescription $R_{dm}=12 [1+0.07(R_0-8~{\rm kpc})]~{\rm kpc}$ and the vertical scale $z_{d}$ is fixed to the best fit value suggested in Ref.~\cite{Freudenreich:1997bx},  $z_{d}$~=~0.340~kpc.

The bulge/bar region is instead characterized by the mass density profile:
\beq
\rho_{bb}(x,y,z)= \bar{\rho}_{bb}\left[ s_a^{-1.85} \,\exp(-s_a) + \exp\left(-\frac{s_b^2}{2}\right) \right] \,
\label{bb}
\eeq
where 
\beq
s_a^2 =  \frac{q_b^2 (x^2+y^2)+z^2}{z_b^2}              
\qquad \quad  {\rm and} \quad \quad 
s_b^4 = \left[ \left(\frac{x}{x_b} \right)^2 + 
\left(\frac{y}{y_b} \right)^2\right]^2 +  \left(\frac{z}{z_b} \right)^4 \,.
\eeq
We implement in this analysis an axisymmetrized version of Eq.~(\ref{bb}), and assume $x_b \simeq y_b = 0.9~{\rm kpc} \cdot (8~{\rm kpc}/R_0)$, $z_b=0.4~{\rm kpc} \cdot (8~{\rm kpc}/R_0)$ and $q_b= 0.6$. See also \cite{Catena:2009mf} concerning the choice of these parameters. Rather than using the two mass normalization scales $\Sigma_{d}$ and $ \bar{\rho}_{bb}$ as free parameters, we re-parameterize these in terms of two dimensionless quantities, namely, the fraction of collapsed baryons $f_{\rm b}$ and the ratio between the bulge/bar and disk masses $\Gamma$:
\beqra
  f_{\rm b} &  \equiv & \frac{\Omega_{\rm DM}+\Omega_{\rm b}}{\Omega_{\rm b}} \frac{M_{bb}+M_d+M_{\textrm{H}_\textrm{I}}+M_{\textrm{H}_2}}{M_{vir}} \label{eq:barpar1} \\
  \Gamma & \equiv & \frac{M_{bb}}{M_d}\,.
  \label{eq:barpar2}
\eeqra
In Eq.~(\ref{eq:barpar1}) we also included the sub-leading contributions to the total virial mass $M_{vir}$ (defined in the following) associated with the atomic ($\textrm{H}_\textrm{I}$) and the molecular ($\textrm{H}_2$) Galactic gas layers, with profiles as given in \cite{dame:267}. In summary, the free parameters describing the luminous components are $R_0$, $R_d$, $f_{b}$ and $\Gamma$.  

Concerning the DM halo component we consider an Einasto profile \cite{Navarro:2003ew,Graham:2006af}, which is favored by the latest N-body simulations and is given by 
\begin{equation}
  \rho_h(r)=\rho^{\prime} f_{E}\left(r/a_h\right)\,,
  \label{nbody}
\end{equation}
with
\begin{equation}
  f_{E}(x) = \exp\left[-\frac{2}{\alpha_E} \left(x^{\alpha_E}-1\right)\right]\,.
\label{eq:einasto}
\end{equation}
The reference normalization $\rho^{\prime}$ and the scale radius $a_h$ in Eq.~(\ref{nbody}) are often rewritten as a function of the virial mass $M_{vir}$ and of the concentration parameter $c_{vir}$ by inverting the relations:
\beqra
  M_{vir} & \equiv & \frac{4\pi}{3} \Delta_{vir} \bar{\rho}_0\,R_{vir}^3 
  = \frac{\Omega_{\rm DM}+\Omega_{\rm b}}{\Omega_{\rm DM}} \,4\pi \int_0^{R_{vir}} dr \, r^2  \rho_h(r) \\
  c_{vir} & \equiv & R_{vir}/r_{-2}
\eeqra
where the virial overdensity $\Delta_{vir}$ in the first equation is computed according to Ref.~\cite{Bryan:1997dn} while $\bar{\rho}_0$ is the mean background density today. The presence in this equation of $\Omega_{\rm DM}$ and $\Omega_{\rm b}$, the DM and baryon energy densities in units of the critical density, reflects our assumption that only a fraction equal to ${\Omega_{\rm DM}}/({\Omega_{\rm DM}+\Omega_{\rm b}})$  of the total virial mass consists of DM. Their values have been set according to the mean values from the fit of the 7-year WMAP data~\cite{Komatsu:2010fb}. In the second equation, instead, $r_{-2}$ is the radius at which the effective logarithmic slope of the DM profile is equal to $-2$. Finally, we assume that the baryons which do not collapse in the disk are distributed according to the same profile of the DM component. The free parameters describing the DM halo are therefore $M_{vir}$, $c_{vir}$ and $\alpha$. 

The Galactic model used to compute $F_h$ also includes an additional parameter, the so-called anisotropy parameter $\beta \equiv 1 - \sigma_{t}^2/\sigma_{r}^2$, where $\sigma_{r}$ and $\sigma_{t}$ are respectively the radial and tangential velocity dispersions of the halo stars. It has been introduced in Ref.~\cite{Catena:2009mf} to include in the parameter estimation the velocity dispersion measurements of Ref.~\cite{Xue:2008se}.

\subsection{XENON1T}
In a real experiment Eq.~(\ref{eq:ddrate2}) has to be modified in order to account for experimental limitations related to the efficiency of the detector, its finite energy resolution and the presence of an experimental noise. In our discussion of the capability of a ton-scale DD experiment in reconstructing the DM mass and SI scattering cross section we will implement the experimental limitations which would apply, as a first approximation, to the XENON1T experiment. We will therefore consider a constant efficiency $\epsilon=0.3$ and include finite energy resolution effects by assuming a Gaussian probability 
\beq
\xi(E,Q) = \frac{1}{\sqrt{2\pi\sigma_{\textrm{Xe}}^2(Q)}} e^{-(E-Q)^2/2\sigma_{\textrm{Xe}}^2(Q)}\,
\label{xi}
\eeq
that to an event of energy $Q$ is wrongly assigned an energy in the interval $E+dE$. In Eq.~(\ref{xi}) the energy dependent dispersion $\sigma_{\textrm{Xe}}(Q)$ is the one estimated for the Xenon detector~\cite{Akrami:2010dn}:
\beq
\sigma_{\textrm{Xe}}(Q) = \sqrt{Q/\textrm{KeV}} (0.579 \,\textrm{KeV}) + 0.021 Q \,. 
\eeq
The observed DD rate is hence equal to
\beq
\frac{dR}{dE} = \epsilon \int_0^\infty dQ \,\xi(E,Q)\,\frac{dR}{dQ}\left(Q\right) \,.
\label{eq:ddrate3}
\eeq  
Regarding the experimental noise, we consider two contributions characterized by a different energy dependence: a flat noise component ({\it i.e.} constant in energy) and an exponential noise component $\propto \exp(-E/E_{0})$ with $E_0=10$ KeV. Finally, in all the conclusions drawn below we will assume an experimental raw exposure of 1000 kg-years and a signal region ranging between 8 KeV and 75 KeV. 

\subsection{Bayesian forecasts}
To test the capability in reconstructing the DM mass and SI cross section of a ton-scale DD experiment with specifications resembling the ones expected for XENON1T, we will first simulate a set of mock DD data and then analyze them employing a Bayesian approach supported by a Markov Chain Monte Carlo (MCMC) Likelihood scanning technique. Mock data have been simulated for each pMSSM benchmark analyzed in this work. 

As already mentioned, in the context of the DD we need to extend the pMSSM parameter space in order to include additional parameters crucial for the computation of the differential rate~(\ref{eq:ddrate2}). These are the parameters introduced in section~\ref{gmod} when defying the Galactic model used to compute the DM phase-space density $F_h$ within Eddington's approach. We denote by ${\bf p}$ the combination of the pMSSM parameters used to describe a certain benchmark and the parameters characterizing the underlying Galactic model. Similarly, by ${\bf p}_b$ we denote the value of ${\bf p}$ at the considered benchmark point. Whereas the assumed pMSSM parameters differ for the four benchmarks studied in this paper, in all cases we adopt the same benchmark Galactic model. This has parameters fixed to the mean values found in \cite{Catena:2011kv}. The mean local DM phase-space density $F_h$ for the Einasto, NFW and Burkert profiles can be downloaded from the online material associated with Ref.~\cite{Catena:2011kv}.

The data simulation consists in sampling a set of $N$ recoil energies $E_i$, $i=1,\dots,N$ from the probability density function $f(E,{\bf p}_b)$ obtained evaluating at the benchmark point ${\bf p}_b$ the spectral function
\beq
f(E,{\bf p}) = \frac{1}{\mu({\bf p})}\left\{ \frac{dR(E,{\bf p})}{dE} + \frac{1}{b-a} + \frac{\exp\left(-E/E_0\right)}{E_{0} \left[\exp(-a/E_0)-\exp(-b/E_0)\right]} \right\} 
\label {sampleE}
\eeq
where $a=$8 KeV, $b=$75 KeV and $\mu({\bf p})=\int_{a}^{b}d\bar{E}\,dR(\bar{E},{\bf p})/dE+2$. This choice corresponds to assume two background events in the signal region during the exposure time:\footnote{As expected for ton-scale DD experiments which aim at reducing the number of background events measured during the exposure time to $\sim1$.} one from the flat noise component and one from the exponential noise component. For a given point ${\bf p}$ in parameter space, Eq.~(\ref{sampleE}) represents the probability that the XENON1T experiment measures an energy (infinitesimally close to) E within the signal region $(a,b)$. The first term on the right hand side of this expression corresponds to the contribution to this probability from the true DM signal while the second and the third terms are associated with the flat and exponential noises respectively. The number of recoil energies $N$ is drawn from a Poisson distribution of mean $\lambda_b$ with probability
\beq
\textrm{Poisson}(N;\lambda_b) = \frac{\lambda_b^N}{N!}\exp\left(-\lambda_b \right)
\eeq
where $\lambda_b\equiv\mu(\bf{p}_b)$ is the number of events ({\it i.e} measured recoil energies) expected for the benchmark under analysis. In Tab.~\ref{tab:spectra2} we report the values of N generated in the simulations analyzed in this work. 

Having explained how to generate mock data from a given benchmark point, we can now focus on their Bayesian analysis. As already mentioned the aim is to reconstruct the dark mater mass and SI cross section. We will estimate the ``preferred value'' for these quantities by determining their marginal posterior probability density functions (PDF). According to Bayes' theorem the posterior PDF of certain parameters ${\bf p}$ is proportional to the product of the Likelihood function $\mathcal{L}(\bf{p,d})$ and the prior probability density $\pi({\bf p})$:
\beq
\mathcal{P}({\bf p,d}) = \frac{\mathcal{L}({\bf d,p}) \pi({\bf p})}{ \mathcal{E}({\bf d})}
\label{pdf}
\eeq
where ${\bf d}$ is the array of datasets used to constrain the parameter space. The Bayesian evidence $\mathcal{E}({\bf d})$, being independent from ${\bf p}$, plays the role of a normalization constant when performing parameter inference. The marginal poster PDF of a generic function $g$ of the parameters ${\bf p}$, {\it e.g.}  $\sigma_{SI}$ and $M_{\chi}$ in our case, is given by the expression
\beq
p(g | {\bf d}) = \int d{\bf p}\,\delta(g({\bf p})-g)\,\mathcal{P}({\bf p}|{\bf d}) \,,
\label{fp}
\eeq
which follows from the definition of conditional probability. MCMC scanning techniques have been then used to sample from Eqs.~(\ref{pdf}) and (\ref{fp}) the marginal posterior PDF of the DM mass a SI cross section.

The Likelihood function is a key ingredient of the present analysis and it has been constructed following the prescriptions given in Ref.~\cite{Akrami:2010dn}. The contribution to the Likelihood from the simulated DD data is given by
\beq
-\log\mathcal{L}_{\textrm{dd}}(E_i,N,{\bf p}) = \mu({\bf p}) - N - N \log\left(\frac{\mu({\bf p})}{N}\right)
-\sum_{i=1}^{N} \log \left( \frac{f(E_i,{\bf p})}{f(E_i,\hat{{\bf p}})} \right)
\eeq
and it has been normalized in such a way that $\log\mathcal{L}_{\textrm{dd}}(E_i,N,\hat{{\bf p}})=0$, where $\mu(\hat{{\bf p}})=N$.
This expression reflects both the information on the number of simulated recoil events $N$, included in the Poissonian pre-factor, and the information about their energies $E_i$, encoded in the spectral function (\ref{sampleE}). In addition to the DD contribution, we have included in the Likelihood a term which assigns a higher statistical weight to points in parameter space characterized by a ``physical mass spectrum''. This term implements in the analysis the LEP2 constraints on SUSY parameters, bounds from collider searches of Higgs decays into $\tau$ pairs (which can be interpreted as an $m_A$-depedent upper bound on the value of $\tan\beta$) and, finally, a lower bound of 600 GeV on the gluino mass \cite{Arbey:2012dq}. These constraints, being expressed in terms of lower limits $\ell_r({\bf d})$ on certain parameters or theoretical predictions $\tau_r({\bf p})$, generate a contribution to the Likelihood function of the following form
\beq
-\log\mathcal{L}^{(r)}_{\textrm{pc}}({\bf d, p}) = \left\{
\begin{array}{c} 
\frac{1}{2}\left( \frac{\tau_r({\bf p})-\ell_r({\bf d}) }{\sigma_r({\bf d})} \right)^2 \qquad \qquad \textrm{if $\tau_r({\bf p})<\ell_r({\bf d})$} \\
\\
0  \qquad \qquad  \qquad \qquad \textrm{otherwise}
\end{array}
\right.
\label{phys}
\eeq
where the index $r$ labels the constraints on the pMSSM mass spectrum previously mentioned and $\sigma_r({\bf d})$ is an estimate of the error associated with the value of the lower limit $\ell_r({\bf d})$. We also included in the Likelihood a term accounting for the recently discovered boson compatible with the Standard Model Higgs:
\beq
-\log\mathcal{L}_{\textrm{Higgs}}({\bf d, p}) = \frac{1}{2}\left( \frac{m_h({\bf p})-\hat{m}_h({\bf d}) }{\sigma_h({\bf d})} \right)^2
\eeq
where $m_{h}({\bf p})$ is the pMSSM prediction for the light CP-even Higgs boson, $\hat{m}_h({\bf d})=126$ GeV and $\sigma_h({\bf d})=3$ GeV. Summing up all the contributions, the final form of the Likelihood function considered in this analysis is
\beq
-\log\mathcal{L}({\bf q},{\bf p})=
-\log\mathcal{L}_{\textrm{dd}}(E_i,N,{\bf p})-\sum_{r}\log\mathcal{L}^{(r)}_{\textrm{pc}}({\bf d},{\bf p}) -\log\mathcal{L}_{\textrm{Higgs}}({\bf d, p}) \,.
\eeq

The last information needed to complete the analysis setup concerns with the choice of the prior PDF $\pi({\bf p})$. In the literature the most investigated choices correspond to flat and logarithmic priors PDF in parameter space. Regarding the pMSSM mass parameters, in the reconstruction of the DM mass and SI cross section we will consider both cases separately to assess the impact of priors on the final results. Concerning instead $\tan\beta$ and the trilinear couplings, following \cite{Bertone:2011nj}, only flat priors will be considered. To carefully sample the posterior PDF of the pMSSM parameters, we will let vary $M_1$, $M_2$, $M_3$, $\mu$, $m_{A}$, $m_{\tilde{q}_{3L}}$, $m_{\tilde{q}_{3R}}$, $m_{\tilde{l}}$ and $m_{\tilde{l}_{3}}$ between 50 GeV and 4 TeV, $A_t$ and $A_b$ between $-$4 TeV and 4 TeV and, finally, $\tan\beta$ between 2 and 60. Galactic model parameters will be treated as nuisance parameters, using the knowledge on their distributions acquired in Refs.~\cite{Catena:2009mf,Catena:2011kv}: the posterior PDF found in these references for the Galactic model parameters through a compilation of various dynamical constraints will be considered in the present analysis as the prior PDF of the astrophysical parameters. As a final remark, we mention here that all the parameters controlling the hadronic matrix elements required to evaluate the DM-nucleus SI scattering cross section have been fixed at the mean values of Ref.~\cite{Akrami:2010dn}. In fact, given the high dimensionality of the pMSSM and the limited constraining power of the data, we preferred to consider in this paper as nuisance parameters only the astrophysical parameters to which we could assign prior PDF motived by previous works. We refer the reader to Refs.~\cite{Cerdeno:2012ix} for an analysis assessing the role of hadronic matrix elements in the context of DM DD.

\subsection{Statistical limitations}
\label{sec:statistical_limitations}
By reconstructing the DM mass and SI cross section we mean what follows: to sample the posterior PDF (\ref{pdf}) through a MCMC algorithm\footnote{In this analysis we employ a modified version of the \supb code interfaced with \ds routines to compute DD observables.} and derive from it the marginal PDF of the DM mass and SI scattering cross section. We will present results in terms of $\alpha\%$ Bayesian credible regions containing therefore a fraction $\alpha$ of the posterior probability. Focusing on credible regions rather than on frequentist confidence intervals allows to circumvent possible issues regarding the coverage of the resulting intervals, a problem which has recently attracted a great attention in the context of the Minimal Supersymmetric Standard Model~\cite{Akrami:2010cz} and in the DM DD field~\cite{Strege:2012kv}.  
\begin{figure}[t]
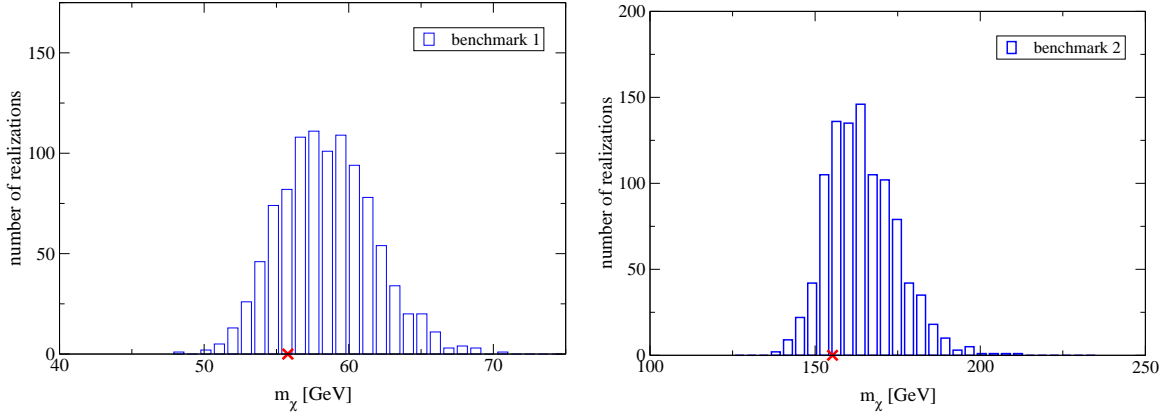

\begin{minipage}[t]{0.49\linewidth}
\centering
\includegraphics[width=\textwidth]{figure/best5.eps}
\end{minipage}
\hspace{0.2cm}
\begin{minipage}[t]{0.49\linewidth}
\centering
\includegraphics[width=\textwidth]{figure/best4.eps}
\end{minipage}
\caption{Histograms showing the distributions of the best fit DM masses derived by fitting 1000 data realizations with the theoretical expectation (\ref{sampleE}). The left panel refers to benchmark {\bf xen1} and the right panel to benchmark {\bf xen2}. The red cross represents the value of the DM mass at the benchmark point.}
\label{fig:xen12}
\end{figure}

There is however another possible source of ``statistical limitations'' affecting the reconstruction procedure described above. This concerns with the fact that the mock data implemented in the analysis, for instance the recoil energies $E_i$, represent just one single realization of the true underlying distribution of the data. If the simulated data do not provide a faithful representation of their true distribution, because of statistical fluctuations, the resulting reconstruction can be significantly biased. In Ref.~\cite{Strege:2012kv} this effect has been observed for DM candidates covering a non negligible spectrum of masses and SI scattering cross sections. It is therefore important to check the sensitivity of our benchmark reconstructions to the particular realization of the data implemented in the analysis. Given the sophistication of the Galactic model of section \ref{gmod} and of the particle physics parameter space considered in this work, to run several MCMC for different realizations of the data is not an efficient way of achieving this task. Therefore, to asses the dependence of our results from the use of a single data sample for each benchmark we followed a different approach based on a simple one dimensional $\chi^2$ minimization: we first sampled for each benchmark a large number of datasets $(N,E_i)$ as explained in the previous subsection. Then, for each realization of the DD data, we fit the resulting distribution of the $N$ energies $E_i$ against the theoretical expectation $(\ref{sampleE})$ varying the DM mass only and fixing the SI scattering cross section and astrophysical parameters to their benchmark values. The resulting distribution of best fit values for the DM mass will allow us to estimate the degree of sensitivity of the benchmarks studied in this paper to the specific data sample considered in the reconstruction of the DM mass and SI cross section. It is in fact the DM mass the quantity mostly affected by statistical fluctuations in the distribution of the simulated recoil energies.
\begin{figure}[t]
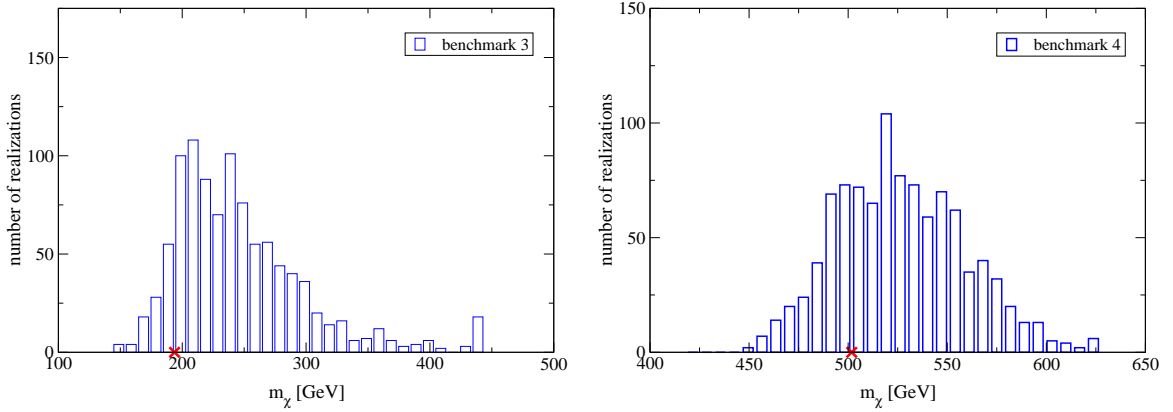

\begin{minipage}[t]{0.49\linewidth}
\centering
\includegraphics[width=\textwidth]{figure/best7.eps}
\end{minipage}
\hspace{0.2cm}
\begin{minipage}[t]{0.49\linewidth}
\centering
\includegraphics[width=\textwidth]{figure/best1.eps}
\end{minipage}
\caption{As in Fig.\ref{fig:xen12} but the left panel refers to benchmark {\bf xen3} and the right panel to benchmark {\bf xen4}.}
\label{fig:xen34}
\end{figure}

The results of this analysis are shown in Fig.\ref{fig:xen12} for benchmarks {\bf xen1} and {\bf xen2}, while Fig.\ref{fig:xen34} refers to benchmarks {\bf xen3} and {\bf xen4}. These histograms show the distributions of the best fit DM masses derived through the fitting procedure explained above for 1000 different data realizations. In each figure the red cross represents the value of the mass at the benchmark point. In all cases we see that the distribution is not perfectly symmetric around the benchmark point: there is a general trend in overestimating the DM mass. This result is in agreement with what found in Ref.~\cite{Strege:2012kv}, where the impact of statistical limitations on the reconstruction of the DM mass and SI cross section has been studied for a broad spectrum of DM candidates. This study shows that the most sensitive DM candidates are those with high masses and low SI cross section, {\it i.e.} candidates associated with a relatively low number of expected DD signal events. This trend is confirmed by the outcome of our analysis, which shows how the benchmarks potentially more sensitive to statistical limitations are those with the highest mass, namely {\bf xen3} and {\bf xen4}, as clearly reported in Fig.\ref{fig:xen34}. The reason behind this phenomenon is that when the number of recoil events is not sufficiently large to produce a faithful description of the underlying recoil energy distribution, a small excess of events at high energy, resulting from a statistical fluctuation, is sufficient to alter the result of the fit: this excess of events at high energy will be better fitted by an artificially large value of DM mass (see also Fig.1 in Ref.~\cite{Strege:2012kv}). 

For the benchmarks studied in this paper we notice however that this effect, though present, it does not influence dramatically the best fit distributions of the DM mass. Since therefore we do not observe any strong pathology related to the selected benchmarks, we continue our analysis focusing on their mass and SI cross section reconstruction at XENON1T. 

\subsection{Mass and cross section reconstruction} 
\begin{figure}[t]
\begin{minipage}[t]{0.48\linewidth}
\centering
\includegraphics[width=\textwidth]{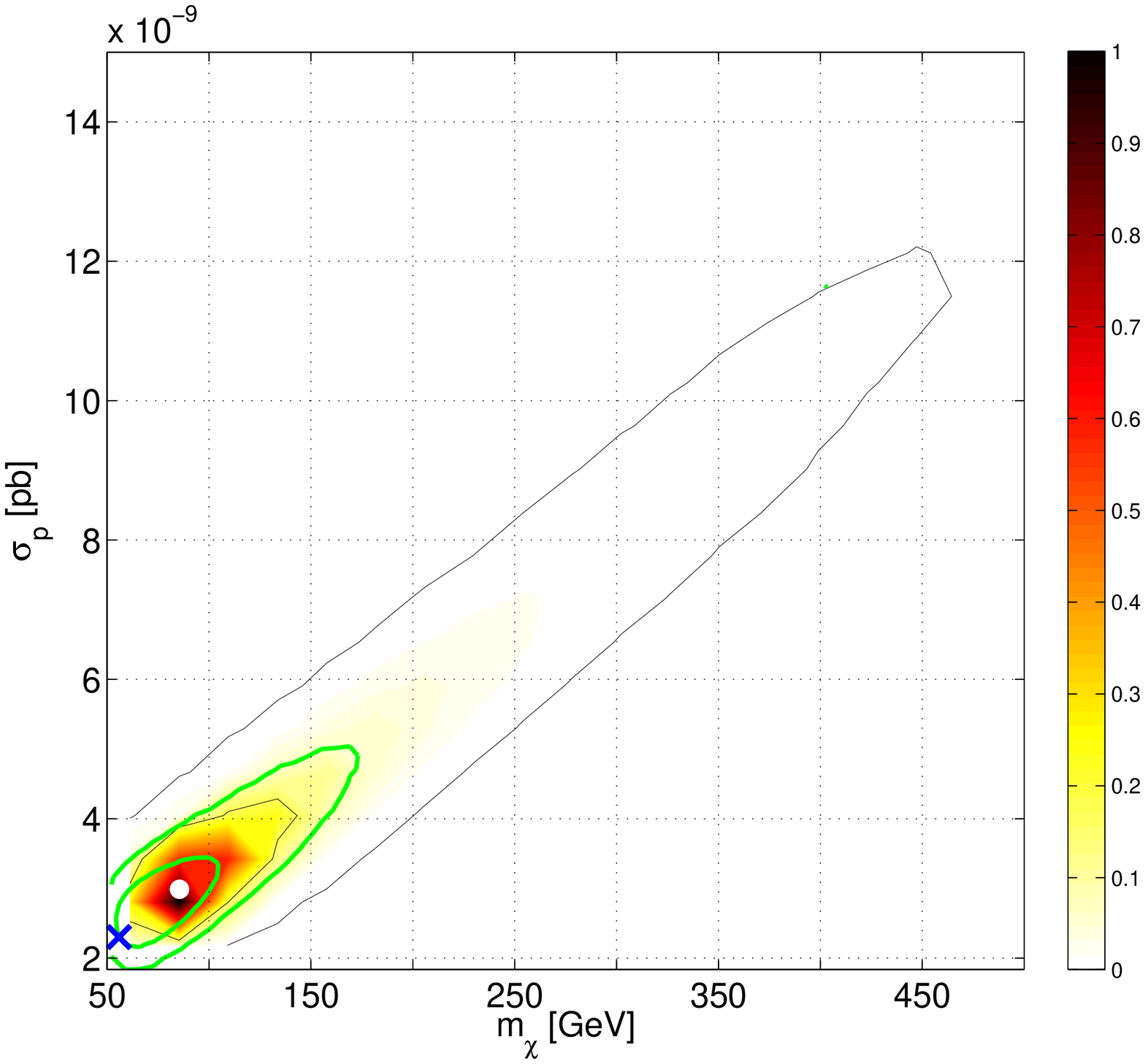}
\end{minipage}
\begin{minipage}[t]{0.48\linewidth}
\centering
\includegraphics[width=\textwidth]{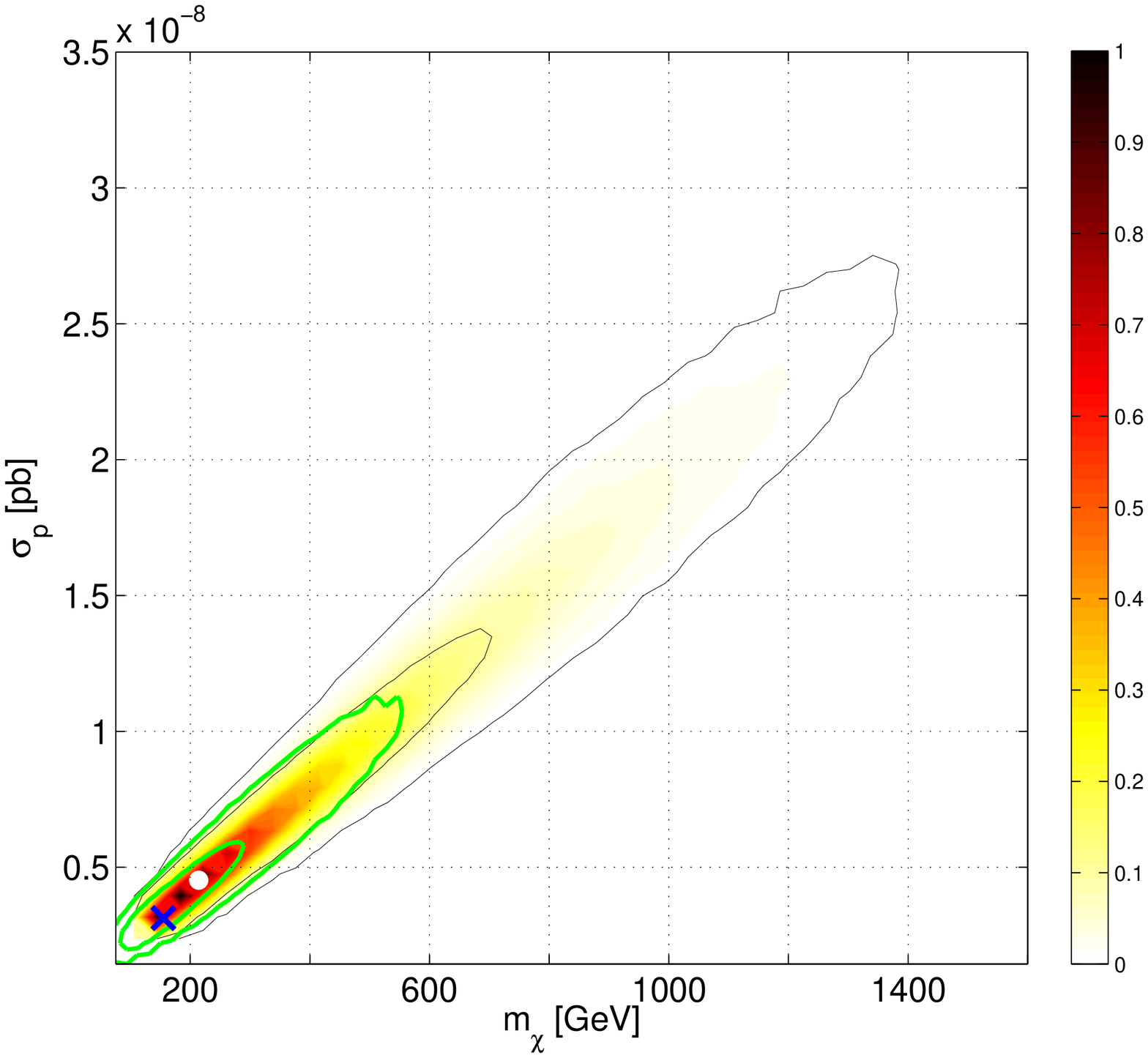}
\end{minipage}
\caption{Two-dimensional marginal posterior PDF in the plane DM mass versus SI scattering cross section. The left panel refers to {\bf xen1} while the right one to {\bf xen2}. In each panel the colored bar follows the posterior PDF, arbitrarily normalized to one, from its peak down to zero. Black contours correspond to the 68$\%$ and 95$\%$ credible regions obtained implementing in the analysis flat priors for the pMSSM mass parameters, while the green contours correspond to the analogous quantities derived from log-priors. The white dot and the blue cross represent respectively the posterior mean (from log-priors) and the benchmark point.}
\label{bm12}
\end{figure}
\begin{figure}[t]
\begin{minipage}[t]{0.48\linewidth}
\centering
\includegraphics[width=\textwidth]{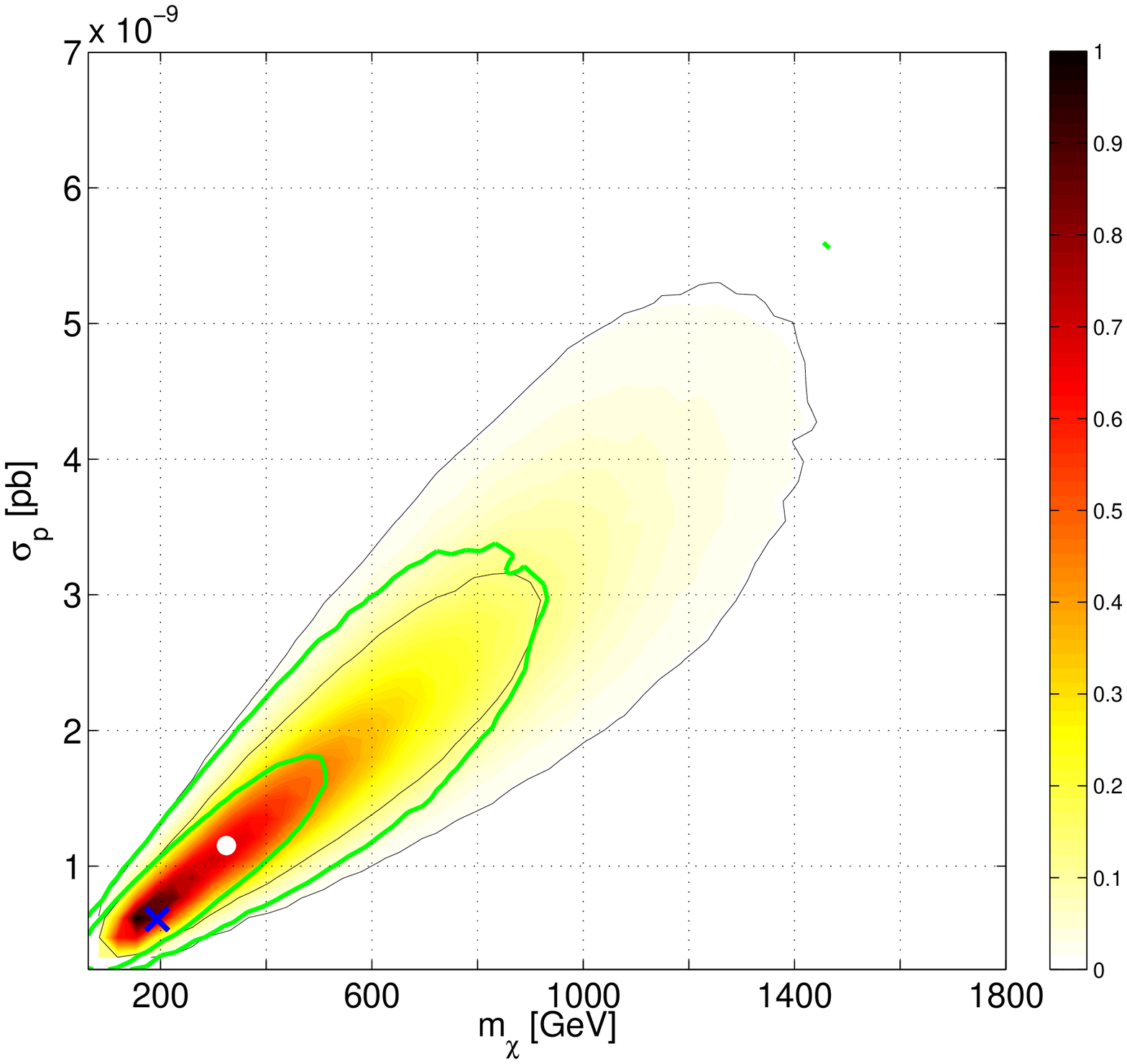}
\end{minipage}
\begin{minipage}[t]{0.48\linewidth}
\centering
\includegraphics[width=\textwidth]{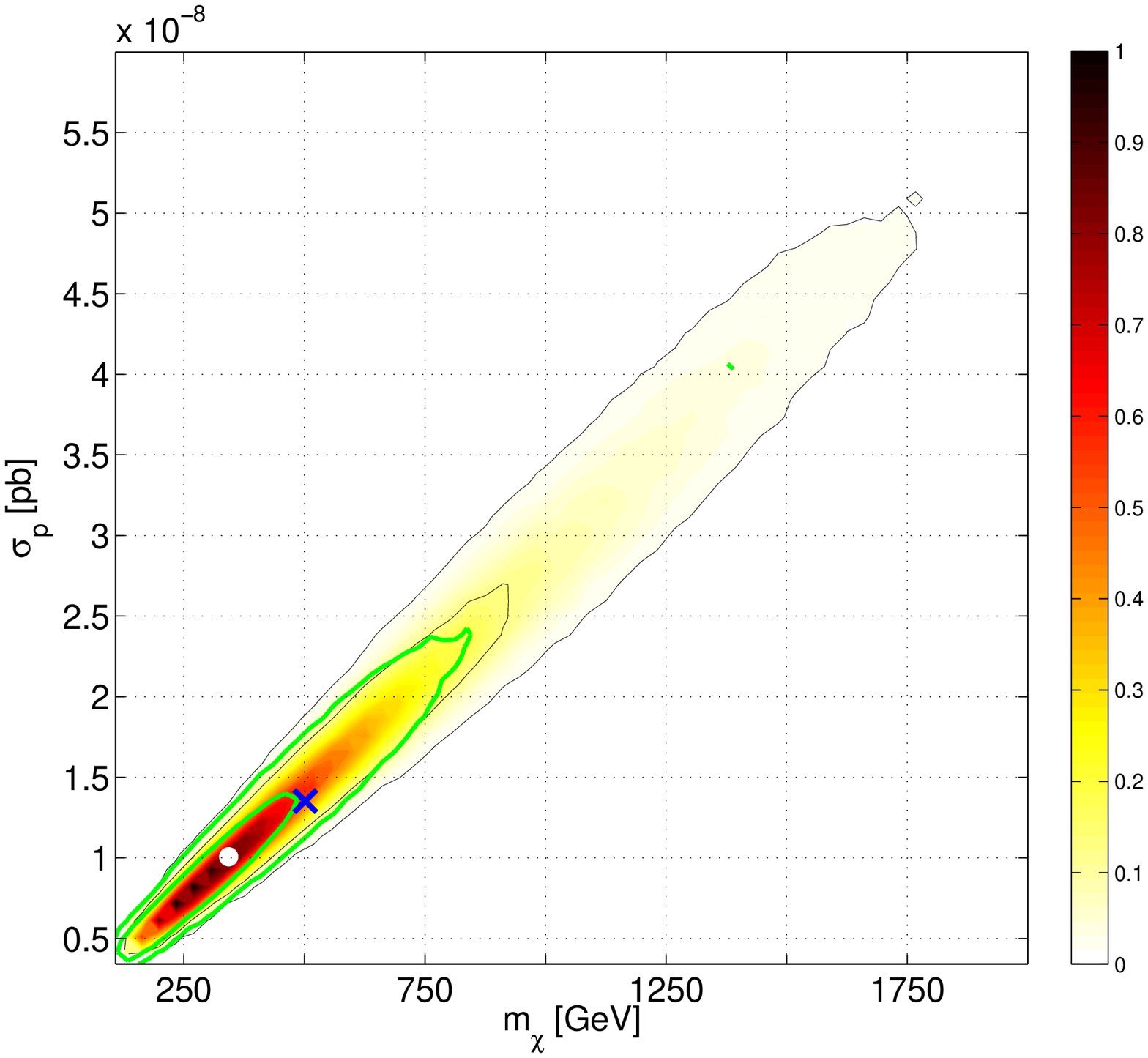}
\end{minipage}
\caption{As in Fig.\ref{bm12} but the left panel refers to benchmark {\bf xen3} and the right panel to benchmark {\bf xen4}.}
\label{bm34}
\end{figure}

The procedure to reconstruct the DM mass and SI cross section introduced in the previous subsections has been applied to the benchmarks {\bf xen1}, {\bf xen2}, {\bf xen3} and {\bf xen4}. For each benchmark we focused on a single data realization considering both flat and log-priors for the pMSSM mass parameters. In Fig.~\ref{bm12} and \ref{bm34} we show for these benchmarks the two-dimensional marginal posterior PDF in the plane DM mass versus SI scattering cross section. In each panel the colored bar follows the two-dimensional posterior PDF which has been conveniently normalized to one. Black contours correspond to the 68$\%$ and 95$\%$ credible regions obtained implementing in the analysis flat priors for the pMSSM mass parameters, while the green contours correspond to the analogous quantities derived from log-priors. Finally, the white dot and the blue cross represent respectively the posterior mean (from log-priors) and the benchmark point. 

The plots confirm the general tendency in overestimating the DM mass described in the previous subsection, here evidenced by the high mass tail featured by the contours and partially by the mismatch between the posterior means and the values of the mass and SI cross section of the original benchmark.  On the other hand we observe a significant improvement of the reconstruction performance when adopting log-priors in the MCMC scanning procedure. Indeed, this choice tends to ``compensate'' possible limitations in the DM mass reconstruction, due to statistical fluctuations in the distribution of the simulated data, by giving more statistical weight to low values of the parameters \cite{Trotta:2008bp} (and therefore to the DM mass), thus partially reducing the high mass tail of the marginal posterior PDF. Because of this compensation mechanism we argue that log-priors provide in this context a more faithful description of the actual potential of upcoming 1-ton DD experiments and we will therefore employ results obtained with this choice when translating our DD analysis into a forecast for the LHC. 

For a given benchmark, in addition to this effect, the quality of the reconstruction is also affected by the value of the DM mass itself. Indeed, the differential recoil rate, for DM masses largely exceeding the one of the nuclei in the target material, depends on the DM parameters only through their ratio $\sigma_{SI}/M_\chi$. Clearly, this degeneracy produces credible regions spanning over ``erroneously'' large portions of the plane $(M_\chi, \sigma_{SI})$. In the present analysis the benchmarks most largely affected by this degeneracy are benchmarks {\bf{xen3}} and {\bf{xen4}}, as also shown in Fig.~\ref{bm34}. 

In the next section we will translate the preferred regions in the plane DM mass versus SI cross section determined here into a prediction for certain classes of signals at the LHC. Since most of the relevant SUSY production cross sections are very steep functions of the pMSSM mass spectrum, and therefore of the DM mass, also the LHC analysis described in the following will be indirectly affected by the statistical limitations producing large uncertainties in the DM mass reconstruction. Large exposures are therefore crucial to enhance the effectiveness of the strategy illustrated here to combine the DD technique with DM searches at the LHC.

In summary, a ton-scale experiment resembling the XENON1T specifications will be able to achieve a robust determination of mass and SI cross section for DM candidates similar to the one associated with the benchmark point {\bf{xen1}}. In fact, in light of our simulations, ${\bf{xen1}}$ clearly emerges as the most promising benchmark considered in this work. This result follows from the large number of recoil events in the simulation of ${\bf{xen1}}$ (170) and from the fact that light candidates are in this context simpler to identify with the DD technique, having a steeper energy spectrum which minimizes the previously mentioned statistical limitations. The benchmark points ${\bf{xen2}}$, ${\bf{xen3}}$ and ${\bf{xen4}}$ feature instead a less efficient reconstruction but still in moderate agreement with the true values of the DM parameters related to the original benchmark. These points are in fact more sensitive to the already mentioned statistical limitations and to the mass/cross section degeneracy, typical of heavy DM candidates.

\section{Collider analysis}
\label{sec:lhc}
Having derived for each benchmark point the accuracy within which the relevant DM properties are reconstructed by a 1-ton DD experiment resembling XENON1T, we now focus on how to translate this information into a prediction for certain LHC observables. Our analysis is organized as follows: first we will reprocess the Markov chains produced in the DD analysis applying to the models in the chains additional constraints from the relic density, limits on the muon anomalous magnetic moment as well as bounds from B-physics processes. Then for each benchmark point we will focus on those LHC processes which are controlled by the same model parameters determining the SI scattering cross section. For these final states we will evaluate the so-called visible cross sections and, when possible, missing energy distributions, concentrating on a few sample points from the chains passing the already mentioned additional constraints and characterized by the highest statistical weight, according to our Bayesian simulation of XENON1T. All the LHC observables will be computed assuming 7 TeV center of mass energy and 4.7 fb$^{-1}$. This procedure is the essence of the method proposed in this paper to correlate the DD technique with DM searches at the LHC.

To rigorously assess the impact on the DD analysis of section \ref{sec:dd} of additional observables one should run new MCMC scans with an extended Likelihood properly incorporating the new constraints. Although this is by now technically achievable in the most simple SUSY realizations (see \cite{Trotta:2008bp,Bertone:2010rv} for some examples) this procedure would be computationally troublesome to implement in the present analysis because of the need of repeated lengthy computations of quantities like the DM relic density, especially in the context of a high dimensional MSSM parameter space, as the one considered in this work, and when studying points in parameter space achieving the correct relict density via coannihilation processes. For these reasons we have employed the less computationally demanding procedure of applying a posteriori the limits from the relic density and flavor processes to the Markov chains generated in the analysis of section \ref{sec:dd}. The limits implemented for $\delta a_\mu$, $b \rightarrow s \gamma$ and $B_s \rightarrow \mu^+ \mu^-$ are the same described in section \ref{sec:pp}. Regarding the relic density, instead, we have just required that it does not exceed the cosmological limit\footnote{Alternatively one could also consider deviations from the standard WIMP paradigm, as enforced by non thermal production \cite{Gelmini:2006mr,Acharya:2009zt,Arcadi:2011ev} or alternative cosmologies \cite{Catena:2004ba,Catena:2007ix}. This would require however a different treatment of the local DM density \cite{Bertone:2010rv}. However, SUSY parameters are less directly correlated in these scenarios.}~\cite{Komatsu:2010fb}. 

The processes considered in this paper are listed in Tab.~\ref{tab:process_summary}. For each of these processes we have simulated the signal possibly detectable by the ATLAS experiment, implement the analysis strategies applied by the ATLAS collaboration \cite{Aad:2009wy} and obtained the number of expected signal events at the luminosities relative to the current available data. The signal events have been generated through the numerical package Madgraph \cite{Alwall:2011uj} combined with the detector simulator PGS normalizing the production cross sections at the NLO  through the package PROSPINO \cite{Beenakker:1996ed}.  The simulations have been validated by generating the main backgrounds for the considered search channels and cross-checking our results with the number of background events inferred by the data. The number of events obtained by the simulations has resulted compatible with expectations.
\begin{table}[t]
\begin{center}
\begin{tabular}{|c|c|c|}
\hline
Search channel & Main processes &  References  \\
\hline
3 leptons + $E_{T}^{\rm miss}$ & $\chi_2^0 \chi_1^{\pm} \rightarrow WZ+\mbox{E}_T^{\rm miss}$ &  \cite{Aad:3l} \\
\hline
2 leptons + $E_{T}^{\rm miss}$ & $\chi_1^{\pm} \chi_2^0$, $\chi_1^{\pm}\chi_1^{\mp}$, $\tilde{l}^{\pm} \tilde{l}^{\mp}$ & \cite{Aad:2l} \\
\hline
monojet + $E_{T}^{\rm miss}$ & $\tilde{t}_1\tilde{t}^{*}_1$  & \cite{ATLAS-CONF-2012-084} \\
\hline 
\end{tabular}
\end{center}
\caption{Summary table of the SUSY searches employed in our collider analysis.}
\label{tab:process_summary}
\end{table}%

\subsection{Analysis of xen1}
\begin{figure}[t]
 \begin{center}
\includegraphics[width=8 cm, height= 7 cm, angle=360]{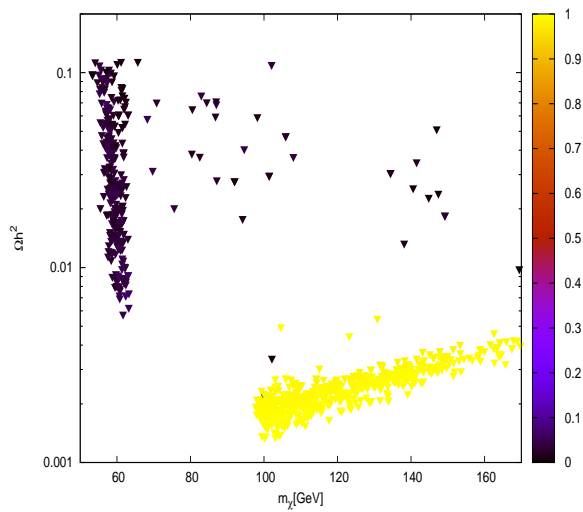}
\end{center}
\caption{Points of the Markov chains associated to the benchmark {\bf{xen1}} passing flavor and relic density constraints in the plane $(M_\chi, \Omega h^2)$ following a color pattern related to the DM higgsino fraction.}
\label{fig:xen1}
\end{figure}

We start our analysis with the lightest DM mass scenario, namely {\bf{xen1}}. Fig.~\ref{fig:xen1}, obtained selecting points in our Markov chains passing flavor and relic density constraints, evidences that values of the DM relic density compatible with current limits can be achieved in two setups only: the first is represented by the two narrow regions lying at values of the DM mass of around 45-60 GeV, thus coinciding with the original benchmark scenario. In this regions a bino-like DM candidate is driven to the correct relic density by resonance enhanced interactions mediated either by the light CP-even Higgs boson or, for slightly lower mass values, by the Z boson. In the second setup, instead, occurring at higher values of the DM mass and corresponding to the yellow region in the plot,  DM is mainly a pure higgsino characterized by very low values of the relic density and by a value of the mass larger than 100 GeV as required by the LEP bound.\footnote{Fig.~\ref{fig:xen1} also reports a very small amount of points at gaugino DM in the mass interval of around 70-300 GeV. For these points a correct value for the relic density is guaranteed by either a low mass slepton or by a CP-odd Higgs. In the first case, however, the slepton should lie very close to the LEP limit and it is thus statistically disfavored by our Monte Carlo scanning procedure, while in the second low values of $m_A$ are disfavored by $Br(B_s \rightarrow \mu^+ \mu^-)$.} 

Notably, there is a clear correspondence between the second setup, represented by the yellow band in Fig.~{\ref{fig:xen1}}, and the high mass tail of the two-dimensional marginal posterior PDF in the plane DM mass versus SI scattering cross section shown in Fig.~\ref{bm12}. As argued in section \ref{sec:statistical_limitations}, this tail arises as a consequence of statistical fluctuations in the recoil energy spectrum affecting the accuracy of the DM mass reconstruction. We therefore conclude that the higgsino DM setup can be effectively disentangled from the bino DM case (corresponding to the original benchmark) by improving the performance of the DD based parameter inference, for instance, by increasing the experimental exposure. In view of this consideration we have limited our collider analysis to a subset of models sampled from the PDF of section \ref{sec:dd} and corresponding to gaugino DM. 

For light DM candidates, as the one considered in this section, the most relevant parameters entering the SI cross section and relic density calculation are the DM mass itself (which is approximately equal to $M_1$), $M_2$ and the value of the $\mu$ parameter. This implies that, from the collider point of view, the most interesting way to probe this scenario consists in the search for direct gaugino production, namely the associated production of $\chi_2^0$ and $\chi_1^{\pm}$, a final state whose properties are also mostly controlled by the same parameters. Assuming that there are no intermediate sleptons between these two states, they can decay either into $W^{\pm}\,Z$, with the bosons which may be also off-shell, or in the pair $Wh$, in addition to missing energy. Our study has been focused on the first possibility employing the ATLAS dedicated search through events with 3 leptons in the final state \cite{Aad:3l}. A strategy for the detection of the $Wh$ channel has been instead proposed in \cite{Baer:2012ts} but it requires much higher luminosities for being detected compared to the previous channel.

\begin{table}[t]
\begin{center}
\begin{tabular}{|c|c|c|c|c|}
\hline
benchmark & $M_2$ & $\mu$ & $\sigma_{\rm vis}[fb]$\, SR1 & $\sigma_{\rm vis}[fb]$\, SR2   \\
\hline
{\bf{xen1-1}} & 498 & 283 & 0.1 & 0.3 \\
\hline
{\bf{xen1-2}} & 215 & 362 & 0.3 & 0.6 \\
\hline
{\bf{xen1-3}} & 212 & 240 & 1.7 & 3.3 \\
\hline
{\bf{xen1-4}} & 138 & 263 & 1.6 & 0.1 \\
\hline
{\bf{xen1-5}} & 266 & 240 & 0.5 & 1.2 \\
\hline
\end{tabular}
\end{center}
\caption{Values of the non-SM visible cross section associated to 3 leptons + $E_T^{\rm miss}$ search channel for 5 sample points relative to the light DM scenario depicted by the benchmark {\bf{xen1}}. The results rely on two signal regions associated, respectively, to the decay of the gauginos into off-shell and on-shell gauge bosons. The experimental limits are $\sigma_{\rm vis}^{\rm obs}=3\,\mbox{fb}$ and $\sigma_{\rm vis}^{\rm obs}=2\,\mbox{fb}$.}
\label{tab:sample_ben5}
\end{table}%

\begin{figure}[t]
\begin{center}
\includegraphics[width=8 cm, height= 7 cm, angle=360]{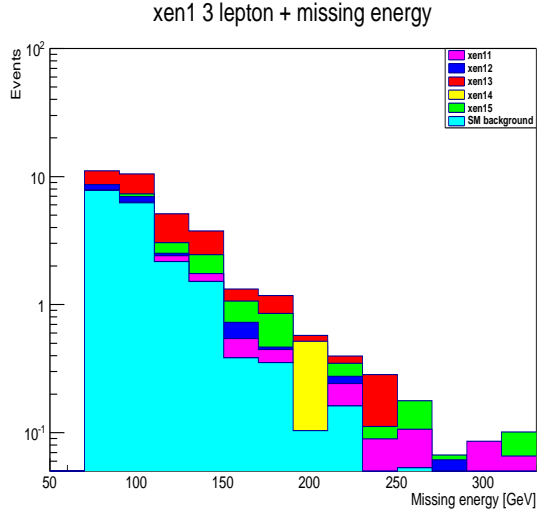}
\end{center}
\caption{Missing energy distribution for the events (signal+background) including three leptons in the final state expected for the models {\bf xen1-1},$\dots$,{\bf xen1-5} sampled from the chains relative to benchmark {\bf{xen1}}.  This result has been obtained in a run of the LHC with an energy in the center of mass of 7 TeV and a luminosity of 4.7 fb$^{-1}$.}
\label{his:3l}
\end{figure}

Tab.~\ref{tab:sample_ben5} reports the results relative to 5 models with high statistical weight, labelled by {\bf xen1-1},$\dots$,{\bf xen1-5}, sampled from the PDF relative to {\bf{xen1}} and passing relic density and flavor constraints. The table reports, for each model, the values of $M_2$ and $\mu$ which determine the characteristics of the produced neutralino and chargino, together with the visible cross sections for non-SM processes, i.e. the product $\sigma\,\epsilon\,A$ of the theoretical production cross section $\sigma$, the detector efficiency $\epsilon$ and the acceptance $A$ which accounts for the reduction of the signal strength due to the analysis cuts. DD combined with relic density and flavor requirements has a good constraining power for the $\mu$ parameter, which we have found to vary on a moderately restricted range with a lower bound of around 200 GeV, as expected from the functional form of the SI cross section, while leaves substantially $M_2$ free to vary. The results in the table shows that the $WZ$ channel can efficiently probe models with moderate values of $M_2$ with some regions already excluded and some other which can be detected in the near future with an $O(1)$ increase of the luminosity. At high values of $M_2$, instead, the chargino and the second neutralino increase their higgsino fraction favoring the decay into $Wh$ and requiring therefore a different search strategy compared to the one considered here involving three leptons in the final state. Fig.~\ref{his:3l}, instead, shows the missing energy distribution corresponding to events (signal$+$background) including three leptons in the final state simulated from the models {\bf xen1-1},$\dots$, {\bf xen1-5} in a run of the LHC with an energy in the center of mass of 7 TeV and a luminosity of 4.7~fb$^{-1}$. 

In summary, for the benchmark scenario {\bf xen1} it is possible to translate the informations on the model parameters derived from our DD simulation into a clearly identifiable signature in the three lepton final state. This is possible because, as already mentioned, both DD and LHC observables crucially depend from the same set of parameters. Remarkably, the results summarized in  Fig.~\ref{his:3l} and Tab.~\ref{tab:sample_ben5} constitute a genuine DD driven forecast for the LHC, successfully obtained by applying our method to the benchmark point {\bf xen1}.

\subsection{Analysis of xen2}
\begin{figure}[t]
\begin{center}
\includegraphics[width=8 cm, height= 7 cm, angle=360]{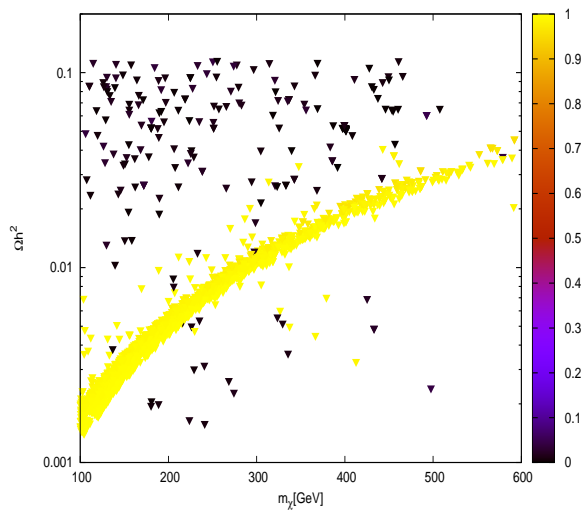}
\end{center}
\caption{Points of the Markov chains associated to the benchmark {\bf{xen2}} passing flavor and relic density constraints in the plane $(M_\chi, \Omega h^2)$ following a color pattern related to the DM higgsino fraction.}
\label{fig:overview_bm4}
\end{figure}

The benchmark point {\bf{xen2}}, similarly to {\bf{xen1}}, has been designed to study MSSM realizations featuring a light EW interacting sector (see e.g.~\cite{Choudhury:2012tc} for a similar approach) evading current constraints which would instead apply to models including colored sectors lying in the multi-TeV region. As for the case of {\bf{xen1}}, in a preliminary investigation of the benchmark {\bf{xen2}}, we selected from our Markov chains the models passing flavor and relic density constraints. These points are shown in Fig.~\ref{fig:overview_bm4} where again the yellow band corresponds to higgsino DM configurations. Contrary to the case of {\bf{xen1}}, however, in the present scenario the higgsino DM models do not only populate the tail of the marginal posterior PDF in the plane DM mass versus SI scattering cross section: comparing in fact Fig.~\ref{bm12} with Fig.~\ref{fig:overview_bm4} one can clearly observe that there are regions in the plane $(M_{\chi},\sigma_{SI})$ with a significant statistical weight corresponding to DM masses above 100 GeV and therefore representing a large fraction of points in the yellow band of Fig.~\ref{fig:overview_bm4}. This is obviously related to the fact that the LEP bound implemented in the Likelihood of our MCMC scans is only effective for exotic states lighter the 100 GeV, while the DM candidate associated with {\bf{xen2}} has a mass of approximately 155 GeV.
Moreover, as consequence of these large uncertainties in the outcome of our DD analysis, the points in our Markov chains passing all the considered bounds, match the current limit on the DM relic density through various mechanisms. The main ones are the coannihilation with a slepton (corresponding to the original benchmark scenario) and resonances occurring in annihilations mediated by the CP-odd Higgs. In addition to these, we have also found configurations in which the parameters $M_1$ and $M_2$ are very close to each other, featuring the already mentioned setup dubbed ``well tempered neutralino''.
For these reasons, the outcome of our DD based analysis cannot resolve alone the degeneracy among pure gaugino and pure higgsino DM configurations (see Fig.~\ref{fig:overview_bm4}). A possible strategy to alleviate this problem could be to encode in the Likelihood the requirement of a DM relic density lying within 2-3 $\sigma$ from the cosmological limit; this requirement cannot be fulfilled by higgsinos with relatively low values of the mass, as required by this benchmark scenario. A potentially interesting alternative is, on the other hand, adding information coming from indirect detection. Indeed bino-like and higgsino DM feature different signatures from the point of view of indirect detection. In particular, current experimental facilities have reached a sensitivity high enough to probe light higgsino DM scenarios with already strong constraints coming form antiproton data \cite{Adriani:2010rc} and new stronger bounds eventually arising in case that the claim of detection of a gamma-ray line in the Fermi-LAT data \cite{Bringmann:2012vr,Weniger:2012tx} will be confirmed (for more details see also e.g.~\cite{Bringmann:2012ez,Buchmuller:2012rc}).  
\begin{table}[t]
\begin{center}
\begin{tabular}{|c|c|c|c|c|c|c|c|}
\hline 
benchmark & $m_\chi$ & $M_2$ & $\mu$ & $m_{\tilde{l}}$ & $m_{\tilde{\tau}}$ &  $\sigma_{\rm vis}[fb] \, SR1$ & $\sigma_{\rm vis}[fb] \, SR2$    \\
\hline
{\bf{xen2-1}} & 132&  348 & 356 & 256 & 139 &  0.13 & 0.09 \\
\hline
{\bf{xen2-2}} & 155&  462 & 940 & 180 & 1100 &  1.24 & 0.20  \\
\hline
{\bf{xen2-3}} & 175&  187 & 554 & 370 & 2500 &  0.03 & 0.02 \\
\hline
{\bf{xen2-4}} & 140&  318 & 388 & 160 & 2300 &  0.18 & 0.11 \\
\hline
{\bf{xen2-5}} & 160 &  3431 & 620 & 2745 & 182 &  0.001 & 0.003  \\
\hline
\end{tabular}
\end{center}
\caption{Values of the non-SM visible cross sections associated to the 3 leptons + $E_T^{\rm miss}$ search channel for 5 sample points relative to the DM scenario depicted by the benchmark {\bf{xen2}}. The results rely on three of the signal regions considered in Ref.~\cite{Aad:3l}. The experimental limits are, respectively, $\sigma_{\rm vis}^{\rm obs}=3\,\mbox{fb}$ and $\sigma_{\rm vis}^{\rm obs}=2\,\mbox{fb}$.}
\label{tab:sample_ben4_3l}
\end{table}%

\begin{table}[t]
\begin{center}
\begin{tabular}{|c|c|c|c|}
\hline
benchmark  &  $\sigma_{\rm vis}[fb] \, SR_{MT2}$ & $\sigma_{\rm vis}[fb] \, SR_{OS}$ & $\sigma_{\rm vis}[fb] \, SR_{SS}$   \\
\hline
{\bf{xen2-1}} &  1.63 & 1.67 & 0.07 \\
\hline
{\bf{xen2-2}} &  0.10 & 0.16 & 0.07 \\
\hline
{\bf{xen2-3}} & 0.39 & 0.57 & 0.04 \\
\hline
{\bf{xen2-4}} & 0.14 & 0.21 & 0.02 \\
\hline
{\bf{xen2-5}} &  0.01 & 0.02 & 0.01 \\
\hline
\end{tabular}
\end{center}
\caption{Values of the non-SM visible cross sections associated to the 2 leptons + $E_T^{\rm miss}$ search channel for 5 sample points relative to the DM scenario depicted by the benchmark {\bf{xen2}}. The results rely on three of the signal regions considered in Ref.~\cite{Aad:2l}. The experimental limits are, respectively, $\sigma_{\rm vis}^{\rm obs}=2\,\mbox{fb}$,  $\sigma_{\rm vis}^{\rm obs}=11.4\,\mbox{fb}$ and $\sigma_{\rm vis}^{\rm obs}=2\,\mbox{fb}$.}
\label{tab:sample_ben4_2l}
\end{table}%

Leaving these possibilities for future works, we will focus in this section on the scenario corresponding to gaugino DM, as an illustrative example of which predictions can be made for the LHC if this benchmark scenario were identified by XENON1T. 
We are therefore neglecting the higgsino DM configurations surviving ({\it i.e.} with a non negligible statistical weight) to our DD analysis and to the additional constraints considered. These configurations would require different LHC search strategies. 
Moreover, we will restrict in the present collider analysis to the DM mass range 135-175 GeV, corresponding to a region of high posterior PDF in the plane DM mass versus scattering cross section able to capture the main features of the present scenario. 

Analogously to the previous benchmark point we still stick on the direct production of EW interacting particles. In addition to three-lepton signals, however, we have also employed detection strategies of events featuring two leptons in the final state \cite{Aad:2l} which improve the sensitivity of the previous search in the case of the presence of sleptons in the decay chain of $\chi_1^{\pm}$ and are, moreover, sensitive to direct slepton production as well as other gaugino production processes like $\chi_1^{\pm}\chi_1^{\mp}$ which can be potentially relevant. As for the case of {\bf xen1}, the logic behind the choice of these final states is that their properties are controlled by the same parameters entering the determination of the SI scattering cross section: in this case, $M_1$, $M_2$ and $\mu$.
 \begin{figure}[t]
\begin{minipage}[htbp]{7.5 cm}
\includegraphics[width=7.5 cm, height= 6.5 cm, angle=360]{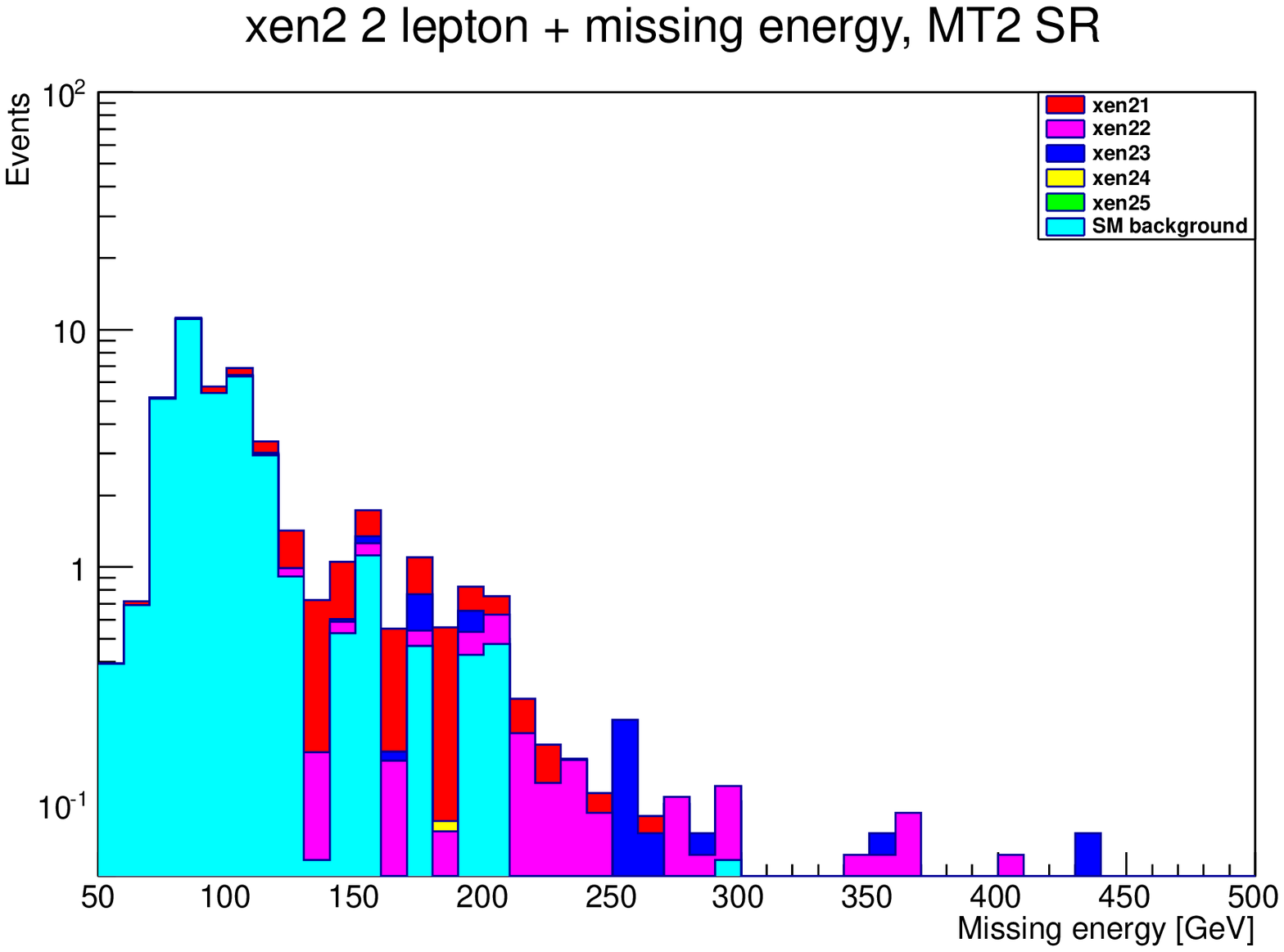}
\end{minipage}
\begin{minipage}[htbp]{7.5 cm}
\includegraphics[width=7.5 cm, height= 6.5 cm, angle=360]{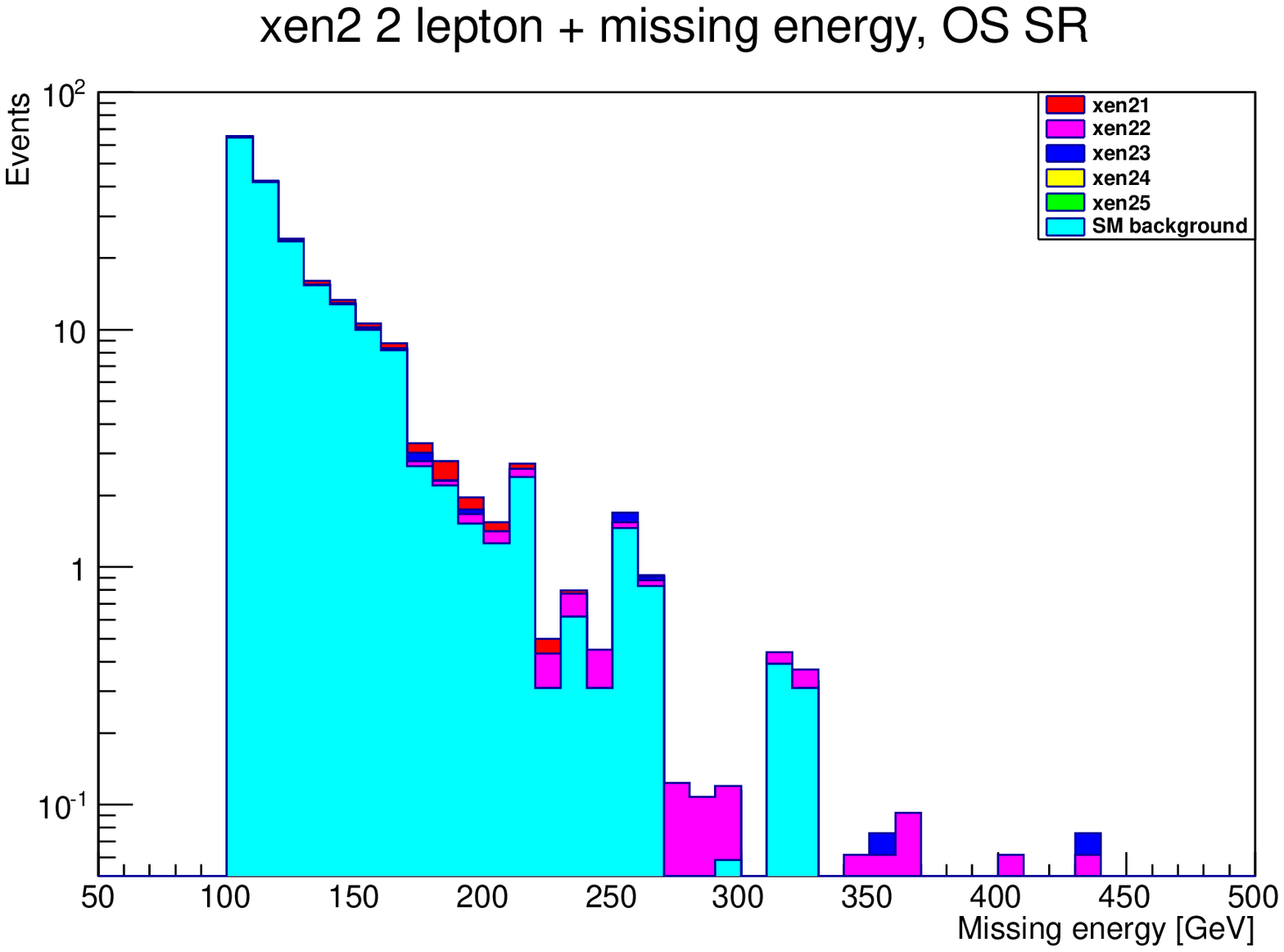}
\end{minipage}
\caption{Missing energy distribution for the events (signal+background) including two leptons in the final state  expected for the models {\bf xen2-1},$\dots$,{\bf xen2-5} sampled from the chains relative to benchmark {\bf{xen2}}.  This result has been obtained in a run of the LHC with an energy in the center of mass of 7 TeV and a luminosity of 4.7 fb$^{-1}$. The left panel refers to the signal region dubbed MT2 while the right panel corresponds to the signal region OS defined in Ref.~\cite{Aad:2l}.}
\label{fig:xen2_2l}
\end{figure}

The outcome of our analysis, performed on 5 representative points (denoted by {\bf xen2-1}, $\dots$,{\bf xen2-5}) sampled from the posterior PDF of {\bf xen2} and reported in Tabs.~(\ref{tab:sample_ben4_3l}) and (\ref{tab:sample_ben4_2l}), shows that the collider prospects are mainly influenced by the value of $M_2$ and by the slepton mass scales while the $\mu$ parameter, partially fixed by DD, is important in determining the composition of $\chi_1^{\pm}$ and $\chi_2^0$. It is also evident that the analysis employed are not sensitive to low mass splittings among the relevant particles. Results of a LHC run simulated from these points focused on final states including two leptons are shown in Fig.~{\ref{fig:xen2_2l}}.

\begin{figure}[t]
\begin{minipage}[htbp]{7 cm}
\includegraphics[width=7 cm, height= 6 cm, angle=360]{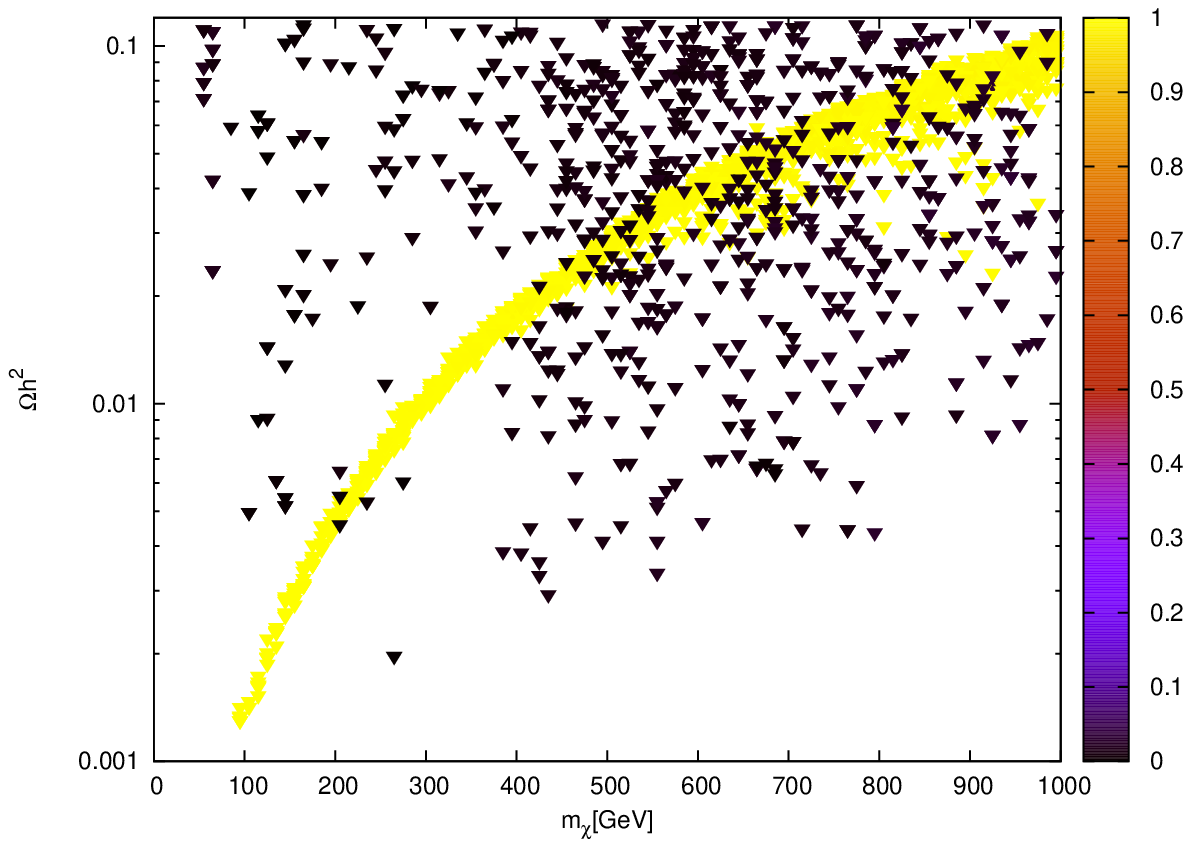}
\end{minipage}
\begin{minipage}[htbp]{7 cm}
\includegraphics[width=7 cm, height= 6 cm, angle=360]{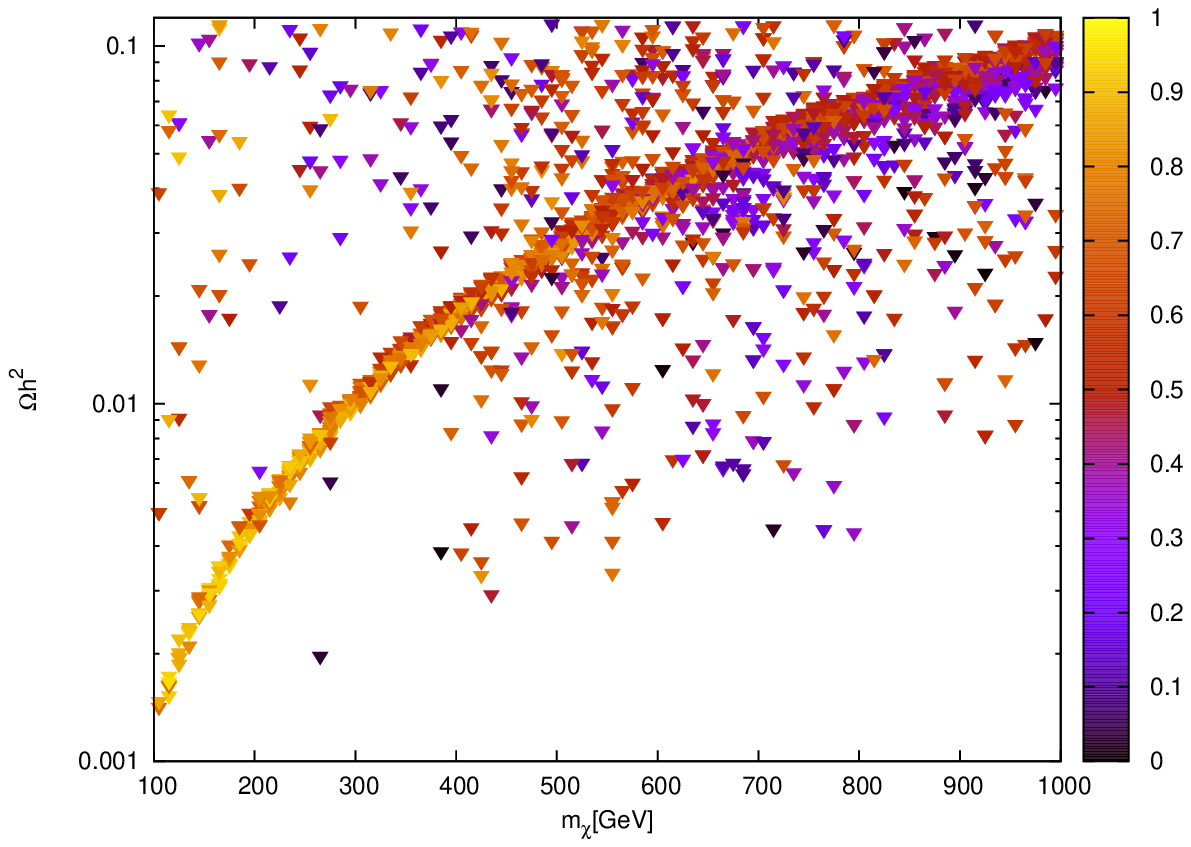}
\end{minipage}\\
\begin{minipage}[htbp]{7 cm}
\includegraphics[width=7 cm, height= 6 cm, angle=360]{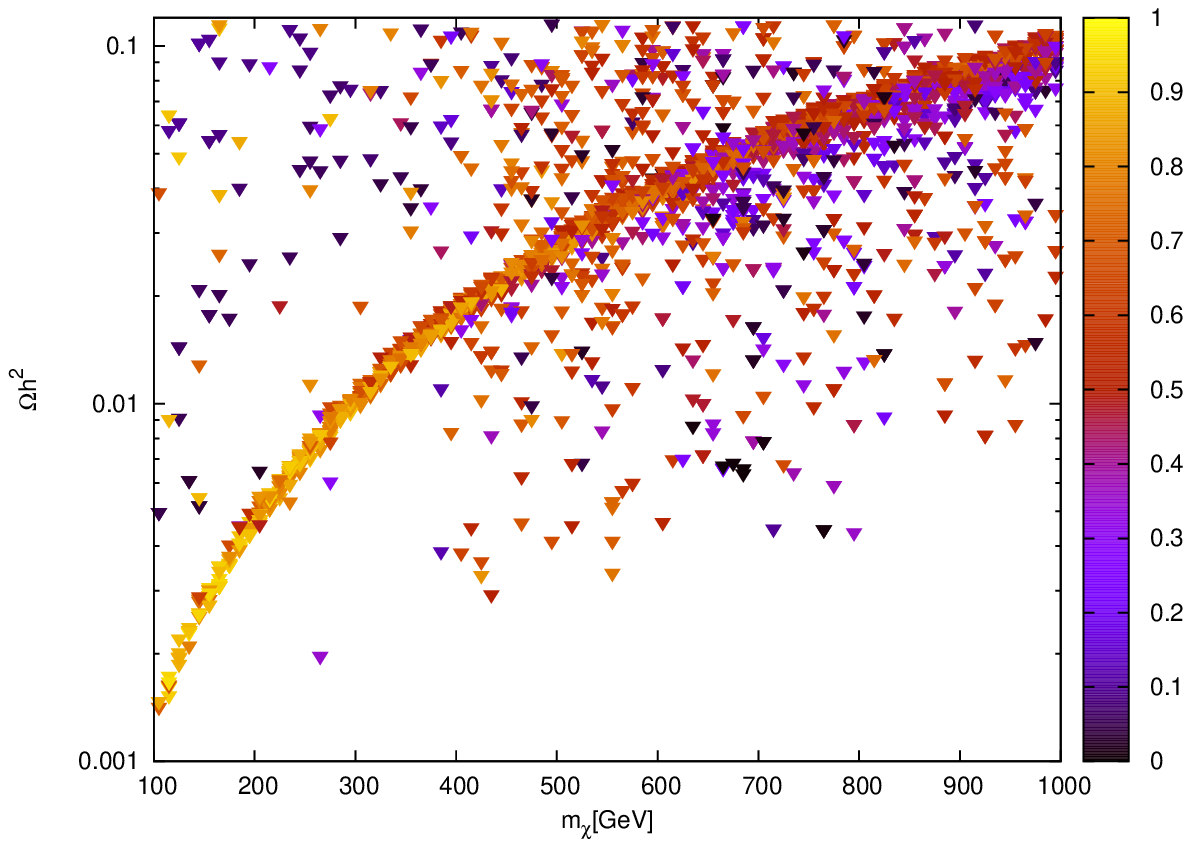}
\end{minipage}
\begin{minipage}[htbp]{7 cm}
\includegraphics[width=7 cm, height= 6 cm, angle=360]{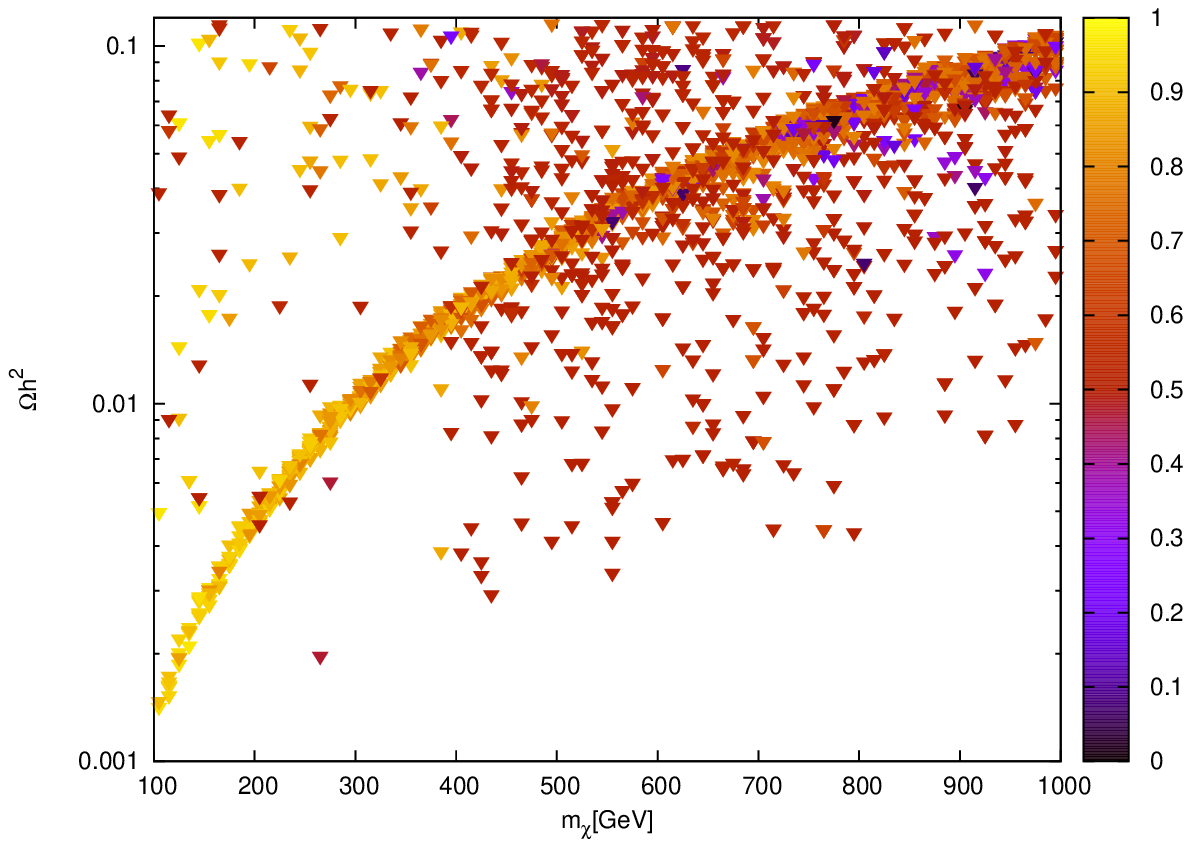}
\end{minipage}
\caption{Points of the Markov chain associated to the benchmark {\bf{xen3}}, selected according the criteria depicted in the text and passing flavor and relic density constraints. In each panel the points are reported in the plane $(m_\chi,\Omega h^2)$. In the upper left panel the points are colored according the corresponding value of the DM higgsino fraction. In the upper right panel instead the color pattern is determined by the quantity $\Delta_A=(m_A-m_\chi)/m_A$. The bottom left and bottom right panels are obtained in analogous manner but the quantities discriminated by the colors are respectively $\Delta_b=(m_{\tilde{b}_1}-m_\chi)/m_{\tilde{b}_{1}}$ and $\Delta_t=(m_{\tilde{t}_1}-m_\chi)/m_{\tilde{t}_{1}}$.}
\label{fig:xen3}
\end{figure}
\subsection{Analysis of xen3}
This benchmark is designed to probe coannihilation scenarios with third family squarks very close to the current experimental limits relative to their direct collider production. As already mentioned, the contribution to the SI scattering cross section of the squark mediated interactions is enhanced for low DM-squark mass splittings and it dominates the cross section itself in the limit of very low higgsino fraction. At the same time these low mass splittings enforce coannihilations driving the DM relic density towards its correct value and imply very simple and definite decay chains at collider. 
\begin{table}[t]
\begin{center}
\begin{tabular}{|c|c|c|c|c|}
\hline
benchmark & $m_{\chi}$ &  $m_{\tilde{t_1}}$ & $m_{\tilde{b_1}}$ &  \\
\hline
{\bf{xen3-1}} & 250 & 524 & 270 & $\tilde{b}_1 \rightarrow b \chi_1^0,\, \tilde{t}_1 \rightarrow t \chi_1^0$ \\
\hline
{\bf{xen3-2}} & 230 & 256 & 442 & $\tilde{b}_1 \rightarrow b \chi_1^0,\, \tilde{t}_1 W^{-} \, \tilde{t}_1 \rightarrow c \chi_1^0$ \\
\hline
{\bf{xen3-3}} & 219 & 249 & 301 &$\tilde{b}_1 \rightarrow b \chi_1^0,\, \tilde{t}_1 \rightarrow c \chi_1^0$ \\
\hline
{\bf{xen3-4}} & 218 & 605 & 234 & $\tilde{b}_1 \rightarrow b\chi_1^0,\, \tilde{t}_1 \rightarrow t \chi_1^0$  \\
\hline
{\bf{xen3-5}} & 198 & 356 & 217 & $\tilde{b}_1 \rightarrow b \chi_1^0,\, \tilde{t}_1 \rightarrow \tilde{b}_1 W^+$ \\ 
\hline
{\bf{xen3-6}} & 150 & 246 & 164 &  $\tilde{b}_1 \rightarrow b \chi_1^0,\, \tilde{t}_1 \rightarrow \tilde{b}_1 W^+$ \\
\hline
{\bf{xen3-7}} & 182 & 540 & 196 & $\tilde{b}_1 \rightarrow b\chi_1^0,\, \tilde{t}_1 \rightarrow t \chi_1^0$ \\
\hline
{\bf{xen3-8}} & 165 & 511 & 175 & $\tilde{b}_1 \rightarrow b\chi_1^0,\, \tilde{t}_1 \rightarrow t \chi_1^0$ \\
\hline
{\bf{xen3-9}} & 172 & 343 & 181 & $\tilde{b}_1 \rightarrow b\chi_1^0,\, \tilde{t}_1 \rightarrow \tilde{b}_1 W^+$  \\
\hline
{\bf{xen3-10}} & 220 & 232 & 380 & $\tilde{b}_1 \rightarrow b \chi_1^0,\, \tilde{t}_1 \rightarrow c \chi_1^0$ \\ 
\hline
\end{tabular}
\end{center}
\caption{Summary table of the points relative to the scenario {\bf{xen3}} used for the collider analysis. The last column reports the main decay channels of the lightest stop and sbottom in the models considered.}
\label{tab:sample_ben7}
\end{table}
As for the other benchmark points, we start the analysis of {\bf xen3} reprocessing the Markov chains produced by the DD simulation applying to the sampled models the current bounds from the DM relic density and flavor physics. The points passing these bounds are shown in Fig.~\ref{fig:xen3}, where the color patterns follow the higgsino fraction (top left panel), $\Delta_A$ (top right panel), $\Delta_b$ (bottom left panel) and, finally  $\Delta_t$ (bottom right panel). The outcome of this analysis shows that extremely pure higgsino DM configurations are favored by the interplay of the DD simulation and the additional constraints subsequently imposed on the resulting Markov chains. Moreover, bino-like DM configurations passing these additional constraints (and corresponding to the original benchmark scenario) achieve the correct relic density mainly through the Higgs funnel mechanism while coannihilations are  disfavored by the requirements on the Higgs mass (a sizable fraction of points in the chains is also excluded by the bounds from $b \rightarrow s \gamma$ and $B_s \rightarrow \mu^+ \mu^-$). This outcome clearly shows that the majority of points in our Markov chains do not correspond to configurations in which the SI scattering cross section is dominated by squark-mediated interactions and thus points again towards the need of combining the information coming from DD with the one coming from other observables, such as for instance the relic density, which constrains more efficiently the coannihilation scenario. 
\begin{table}[t]
\begin{center}
\begin{tabular}{|c|c|c|c|c|c|}
\hline
benchmark & SR1 & SR2 & SR3 & SR4   \\
\hline
$\sigma_{\rm vis}^{\rm obs}[\mbox{pb}]$ & 1.63 & 0.13 & 0.026 & 0.006  \\
\hline
{\bf{xen3-1}} & 0.2 & 0.04 & 0.005 & 0.0008  \\
\hline
{\bf{xen3-2}} &  0.25 & 0.04 & 0.005 & 0.001\\
\hline
{\bf{xen3-3}} & 0.3 & 0.05 & 0.007 & 0.001  \\
\hline
{\bf{xen3-4}} & 0.4 & 0.07 & 0.010 & 0.0015 \\
\hline
{\bf{xen3-5}} & 0.5 & 0.08 & 0.010 & 0.001  \\ 
\hline
{\bf{xen3-6}} & 1.59 & 0.19 & 0.025 & 0.002  \\
\hline
{\bf{xen3-7}} &  0.98 & 0.13 & 0.019 & 0.002\\
\hline
{\bf{xen3-8}} & 1.66 & 0.23 & 0.035 & 0.007  \\
\hline
{\bf{xen3-9}} & 1.37 & 0.18 & 0.027 & 0.004 \\
\hline
{\bf{xen3-10}} & 0.51 & 0.08 & 0.011 & 0.002  \\ 
\hline
\end{tabular}
\end{center}
\caption{Values of the non-SM visible cross sections of the points used for the collider analysis with respect to the 4 signal regions defined in Ref.~\cite{ATLAS-CONF-2012-084}. This analysis is aimed at exploring processes leading to sensitively soft jets and then is particularly suitable to probe coannihilation scenarios.}
\label{tab:monojet_ben7}
\end{table}%

In spite of these difficulties found in the DD analysis of the coannihilation benchmark {\bf xen3}, we nevertheless tried to extract as many information as possible from this result, translating the associated limits on the model parameters into certain predictions for the LHC.  
Analogously to the case of {\bf xen2}, we will restrict to bino-like configurations emerging from our Monte Carlo scans, considering only regions of the ($M_{\chi},\sigma_{SI}$) plane exhibiting a high statistical weight and capturing all the features relevant for the present analysis, namely, a portion of this plane corresponding to DM masses lower than 250 GeV.
As for the other benchmark points, we now focus on a final state whose properties are determined by the same parameters entering in the calculation of the SI scattering cross section and then for this process we evaluate LHC observables quantities, such as for instance visible cross sections. For this benchmark point a process with these properties is the a final state involving pairs of stop and sbottom squarks. If a coannihilating sbottom is the NLSP, this can only decay through the channel  $\tilde{b} \rightarrow b \chi_1^0$ while in the case of a stop NLSP the dominant decay channel is often $t \rightarrow c \chi_1^0$, arising at one-loop. Another possible decay is $\tilde{t} \rightarrow b \overline{f} f^{\prime} \chi_1^0$, with $f$ and $f^{\prime}$ being two light fermions. These decays, however, evade most of the current SUSY searches since their products are too soft to pass the analysis cuts on the $p_T$, or even the requirements for correct reconstruction of the final states. Such configurations thus require search strategies optimized for very low mass splittings or alternatively can be probed using initial state radiation. In this work we will concentrate on monojet searches and employ the dedicated ATLAS analysis strategy of Ref.~\cite{Aad:2012ky}. Good prospects of probing scenarios with very low mass splitting have been obtained as well from the study of monophoton events in Ref.~\cite{Belanger:2012mk}.  

To this purpose we have sampled from the PDF obtained from our DD analysis  a few representative models (listed in Tab.~\ref{tab:sample_ben7}) and generated from them LHC final states involving pairs of stop and sbottom squarks with an additional jet. The results, reported in  Tab.~\ref{tab:monojet_ben7}, show that this coannihilation scenario is rather efficiently probed, with some points already excluded, especially at low masses while we have a significant decrease in the performance at higher values since the stop/sbottom production cross sections are rather steeply decreasing functions of the masses of these particles. These results further stress the importance of increasing the performance in reconstructing the DM mass (in this case related by coannihilation requirements to the squark ones) since even slightly different determinations drastically alter the collider detection prospects.  

\subsection{Analysis of xen4}
\begin{figure}[t]
\begin{minipage}[htbp]{7.5 cm}
\includegraphics[width=7.5 cm, height= 6.5 cm, angle=360]{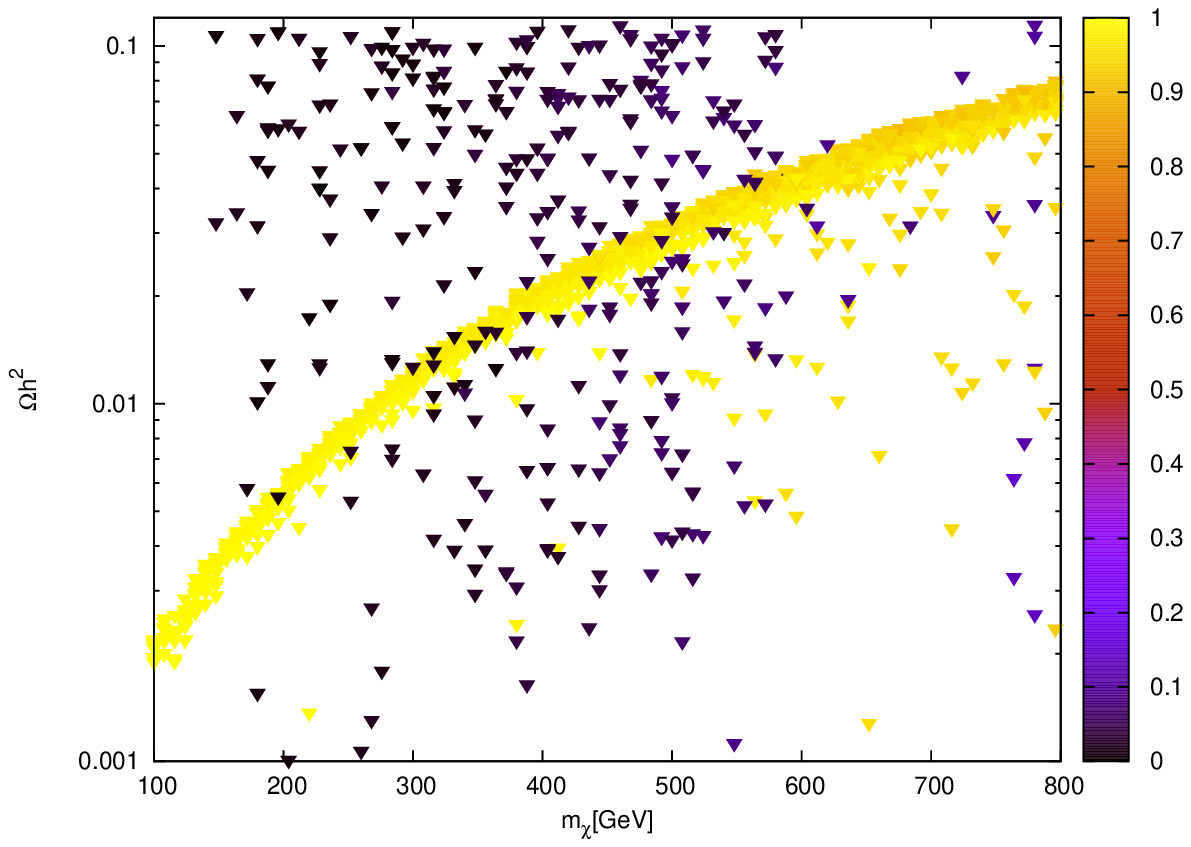}
\end{minipage}
\begin{minipage}[htbp]{7.5 cm}
\includegraphics[width=7.5 cm, height= 6.5 cm, angle=360]{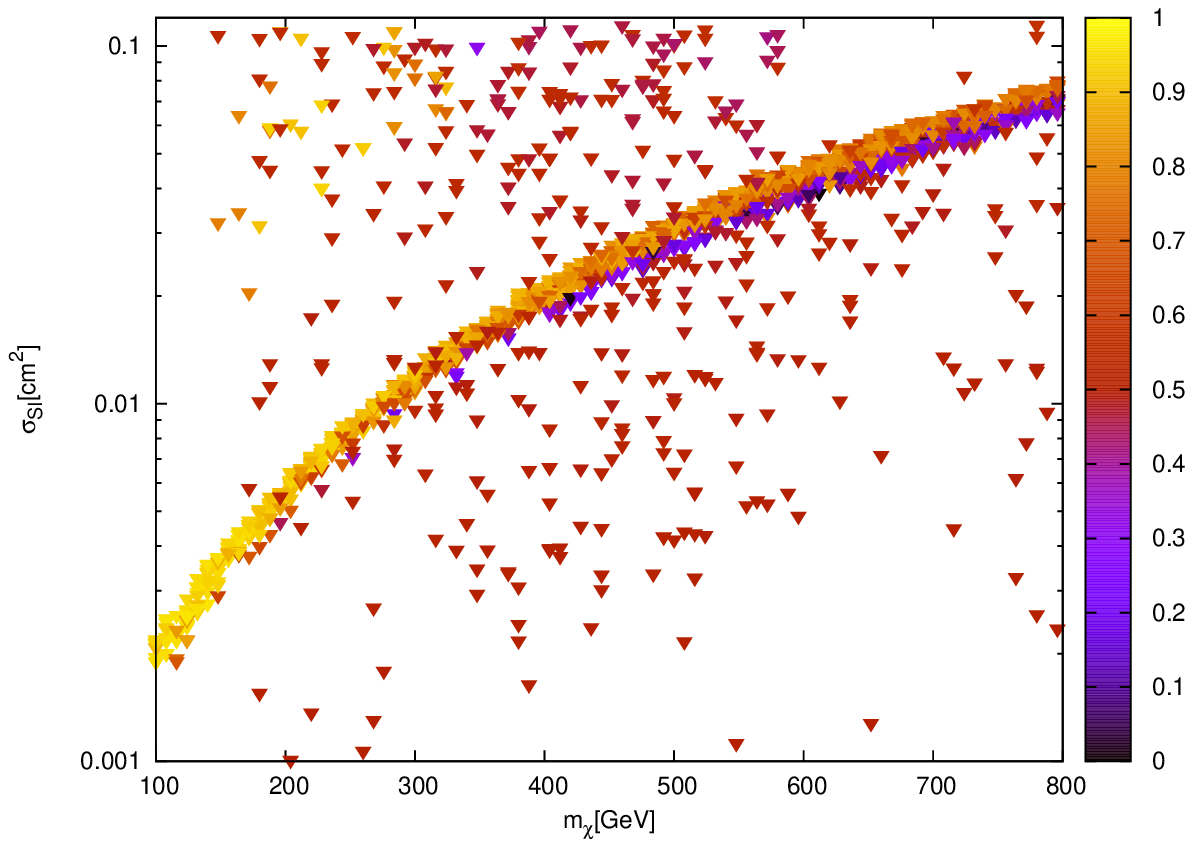}
\end{minipage}
\caption{Points in the Markov chain associated to the benchmark {\bf{xen4}} passing flavor and relic density constraints. The points in the two plots follow color patterns determined, respectively, by higgsino fraction and $\Delta_A$. By comparing the two panels one notices that a large fraction of points at gaugino DM clusters around the value $\Delta_A\simeq 0.5$.}
\label{fig:xen4}
\end{figure}

The last benchmark point considered in the analysis has been also designed, similarly to {\bf xen3}, to investigate a scenario where DM is produced through coanniliations involving squarks of the third family. In this case, however, the DM mass is significantly higher than in the case of the previous benchmark. Fig.~\ref{fig:xen4} clearly shows that no coannihilation configurations emerged from the interplay of the DD simulations and the application of the relic density and flavor bounds, being the PDF samples strongly dominated by higgsino DM candidates and, to some extent, by CP-odd Higgs funnel configurations. In view of this outcome, we regard the reconstruction of this point as unsuccessful and thus we will not pursue any collider analysis for this scenario. 

\section{Conclusions}
\label{sec:conclusions}
In this paper we have explored a new strategy to combine direct and collider DM searches. The idea is very simple: the DD technique is directly sensitive to the DM mass and scattering cross section only. The same technique, however, has also the potential to indirectly constrain the model parameters (or certain combination of them) entering the calculation of the relevant DD scattering cross sections. Therefore, focusing on LHC processes whose properties are also determined by the very same model parameters, an hypothetically identified DD signal can be directly translated into a genuine DD driven forecast for the missing energy distributions associated with these processes.

This analysis has been performed in the context of the so-called phenomenological Minimal Supersymmetric Standard Model where a large number of possible supersymmetric configurations is conveniently described by a limited number of parameters. We focused on four benchmark points representative of WIMP DM candidates which should be soon probed by the upcoming generation of 1-ton DD experiments and corresponding to different DM production scenarios. In all cases a clear correlation pattern between DD and LHC observables is enforced by the underlying mechanism accounting for the DM relic density.

For each benchmark, we tested the capability of a 1-ton DD experiment resembling the XENON1T experiment in reconstructing the DM mass and SI scattering cross section within the framework of a Bayesian analysis supported by a Markov Chain Monte Carlo scanning technique. In our analysis we carefully addressed the impact on the reconstructions of the astrophysical uncertainties entering the computation of the differential count rates, of possible statistical limitations related to the finite number of measured recoil energies as well as of the choice of the priors for the model parameters. 

We derived through this procedure the accuracy within which the DM mass and scattering cross section are reconstructed by the Xenon experiment in its 1-ton configuration. We then focused on the regions of the plane DM mass versus scattering cross section enclosing the 95$\%$ of the posterior probability. Within these regions we selected a subset of models by reprocessing our Markov chains introducing additional constraints from the relic density, the anomalous magnetic moment of the muon and B-physics processes. For some of these models we simulated the ATLAS detector response in its configuration with 7 TeV center of mass energy and carefully studied the prospects of detecting certain classes of final states for which dedicated search strategies have been developed by the ATLAS collaboration and, most importantly, whose properties crucially depend from the same model parameters entering the calculation of the SI scattering cross section associated with the corresponding benchmark point. 

The procedure proposed here defines a method to efficiently exploit the upcoming generations of 1-ton DD experiments to forecast valuable predictions for the LHC and it constitutes the main achievement of this paper. The results obtained within this framework are particularly promising for the benchmark ${\bf xen1}$ which is representative of a light bino-like neutralino produced via resonant annihilations. In this case a definite prediction for a LHC final state including three leptons and missing energy can be elaborated starting from the informations derived from the simulated 1-ton DD experiment. The other benchmarks, instead, are more sensitive to the DD statistical limitations discussed in the text and affecting the accuracy of the DM mass reconstruction. Larger exposures in the DD (here we considered 1000 kg-years) and eventually more informative Likelihoods can however further enhance the effectiveness of our approach and allow to draw interesting conclusions also in the case of challenging DM candidates as for instance ${\bf xen 4}$.

In summary, we have proposed a strategy to systematically translate an hypothetical DD signal into a prediction for the LHC. This method has the potential to significantly strengthen the complementarity between the DD technique and DM searches at collider and will be further investigated in future works.

\acknowledgments
It is a pleasure to thank Antonio Masiero for his collaboration during the early stages of the project and useful discussions on the main idea behind this work. We also thank Umberto de Sanctis, Alberto Tonero, Michele Pinamonti, Andrea De Simone and Florian Bonnet for fruitful conversations. These results have been partially obtained making use of the bwGRiD Cluster (http://www.bw-grid.de), member of the German D-Grid initiative, funded by the Ministry for Education and Research (Bundesministerium f\"ur Bildung und Forschung) and the Ministry for Science, Research and Arts Baden-W\"urttemberg (Ministerium f\"ur Wissenschaft, Forschung und Kunst Baden-W\"urttemberg).
G.A and P.U. acknowledges partial support from the European Union FP7  ITN-INVISIBLES (Marie Curie Actions, PITN- GA-2011- 289442). G.A also thanks the Galileo Galilei Institute for the warm hospitality during part of the realization of this work.

\bibliography{collider}{}

\begin{thebibliography}{10}

\bibitem{Fox:2011pm}
P.~J. Fox, R.~Harnik, J.~Kopp, and Y.~Tsai, ``{Missing Energy Signatures of
  Dark Matter at the LHC},'' {\em Phys.Rev.}, vol.~D85, p.~056011, 2012,
  1109.4398.

\bibitem{xenon:2012nq}
 {\em et~al.}, ``{Dark Matter Results from 225 Live Days of XENON100 Data},''
  2012, 1207.5988.

\bibitem{Catena:2009mf}
R.~Catena and P.~Ullio, ``{A novel determination of the local dark matter
  density},'' {\em JCAP}, vol.~1008, p.~004, 2010, 0907.0018.

\bibitem{Catena:2011kv}
R.~Catena and P.~Ullio, ``{The local dark matter phase-space density and impact
  on WIMP direct detection},'' {\em JCAP}, vol.~1205, p.~005, 2012, 1111.3556.

\bibitem{Iocco:2011jz}
F.~Iocco, M.~Pato, G.~Bertone, and P.~Jetzer, ``{Dark Matter distribution in
  the Milky Way: microlensing and dynamical constraints},'' {\em JCAP},
  vol.~1111, p.~029, 2011, 1107.5810.

\bibitem{Garbari:2011dh}
S.~Garbari, J.~I. Read, and G.~Lake, ``{Limits on the local dark matter
  density},'' {\em Mon.Not.Roy.Astron.Soc.}, vol.~416, pp.~2318--2340, 2011,
  1105.6339.

\bibitem{Pato:2010yq}
M.~Pato, O.~Agertz, G.~Bertone, B.~Moore, and R.~Teyssier, ``{Systematic
  uncertainties in the determination of the local dark matter density},'' {\em
  Phys.Rev.}, vol.~D82, p.~023531, 2010, 1006.1322.

\bibitem{Salucci:2010qr}
P.~Salucci, F.~Nesti, G.~Gentile, and C.~Martins, ``{The dark matter density at
  the Sun's location},'' {\em Astron.Astrophys.}, vol.~523, p.~A83, 2010,
  1003.3101.

\bibitem{deBoer:2010eh}
W.~de~Boer and M.~Weber, ``{The Dark Matter Density in the Solar Neighborhood
  reconsidered},'' {\em JCAP}, vol.~1104, p.~002, 2011, 1011.6323.

\bibitem{Hansen:2005yj}
S.~H. Hansen, B.~Moore, M.~Zemp, and J.~Stadel, ``{A Universal velocity
  distribution of relaxed collisionless structures},'' {\em JCAP}, vol.~0601,
  p.~014, 2006, astro-ph/0505420.

\bibitem{Strege:2012kv}
C.~Strege, R.~Trotta, G.~Bertone, A.~H. Peter, and P.~Scott, ``{Fundamental
  statistical limitations of future dark matter direct detection
  experiments},'' {\em Phys.Rev.}, vol.~D86, p.~023507, 2012, 1201.3631.

\bibitem{Djouadi:1998di}
A.~Djouadi {\em et~al.}, ``{The Minimal supersymmetric standard model: Group
  summary report},'' 1998, hep-ph/9901246.

\bibitem{Aad:2011ib}
G.~Aad {\em et~al.}, ``{Search for squarks and gluinos using final states with
  jets and missing transverse momentum with the ATLAS detector in sqrt(s) = 7
  TeV proton-proton collisions},'' 2011, 1109.6572.

\bibitem{Chatrchyan:2011zy}
S.~Chatrchyan {\em et~al.}, ``{Search for Supersymmetry at the LHC in Events
  with Jets and Missing Transverse Energy},'' {\em Phys.Rev.Lett.}, vol.~107,
  p.~221804, 2011, 1109.2352.

\bibitem{atlas:2012gk}
G.~Aad {\em et~al.}, ``{Observation of a new particle in the search for the
  Standard Model Higgs boson with the ATLAS detector at the LHC},'' {\em
  Phys.Lett.}, vol.~B716, pp.~1--29, 2012, 1207.7214.

\bibitem{Giardino:2012dp}
P.~P. Giardino, K.~Kannike, M.~Raidal, and A.~Strumia, ``{Is the resonance at
  125 GeV the Higgs boson?},'' 2012, 1207.1347.

\bibitem{Carmi:2012in}
D.~Carmi, A.~Falkowski, E.~Kuflik, T.~Volansky, and J.~Zupan, ``{Higgs After
  the Discovery: A Status Report},'' 2012, 1207.1718.

\bibitem{Haber:1995be}
H.~E. Haber, ``{Challenges for nonminimal Higgs searches at future
  colliders},'' 1995, hep-ph/9505240.

\bibitem{Baer:2011ab}
H.~Baer, V.~Barger, and A.~Mustafayev, ``{Implications of a 125 GeV Higgs
  scalar for LHC SUSY and neutralino dark matter searches},'' {\em Phys.Rev.},
  vol.~D85, p.~075010, 2012, 1112.3017.

\bibitem{Arbey:2011ab}
A.~Arbey, M.~Battaglia, A.~Djouadi, F.~Mahmoudi, and J.~Quevillon,
  ``{Implications of a 125 GeV Higgs for supersymmetric models},'' {\em
  Phys.Lett.}, vol.~B708, pp.~162--169, 2012, 1112.3028.

\bibitem{Hall:2011aa}
L.~J. Hall, D.~Pinner, and J.~T. Ruderman, ``{A Natural SUSY Higgs Near 126
  GeV},'' {\em JHEP}, vol.~1204, p.~131, 2012, 1112.2703.

\bibitem{Arbey:2012dq}
A.~Arbey, M.~Battaglia, A.~Djouadi, and F.~Mahmoudi, ``{The Higgs sector of the
  phenomenological MSSM in the light of the Higgs boson discovery},'' 2012,
  1207.1348.

\bibitem{Aad:2011rv}
G.~Aad {\em et~al.}, ``{Search for neutral MSSM Higgs bosons decaying to
  $\tau^+ \tau^-$ pairs in proton-proton collisions at sqrt(s) = 7 TeV with the
  ATLAS detector},'' {\em Phys.Lett.}, vol.~B705, pp.~174--192, 2011,
  1107.5003.

\bibitem{Chatrchyan:2012vp}
S.~Chatrchyan {\em et~al.}, ``{Search for neutral Higgs bosons decaying to tau
  pairs in pp collisions at sqrt(s)=7 TeV},'' {\em Phys.Lett.}, vol.~B713,
  pp.~68--90, 2012, 1202.4083.

\bibitem{Giardino:2012ww}
P.~P. Giardino, K.~Kannike, M.~Raidal, and A.~Strumia, ``{Reconstructing Higgs
  boson properties from the LHC and Tevatron data},'' {\em JHEP}, vol.~1206,
  p.~117, 2012, 1203.4254.

\bibitem{Moroi:1995yh}
T.~Moroi, ``{The Muon anomalous magnetic dipole moment in the minimal
  supersymmetric standard model},'' {\em Phys.Rev.}, vol.~D53, pp.~6565--6575,
  1996, hep-ph/9512396.

\bibitem{Calibbi:2011dn}
L.~Calibbi, R.~Hodgkinson, J.~Jones~Perez, A.~Masiero, and O.~Vives, ``{Flavour
  and Collider Interplay for SUSY at LHC7},'' {\em Eur.Phys.J.}, vol.~C72,
  p.~1863, 2012, 1111.0176.

\bibitem{Lunghi:2006hc}
E.~Lunghi and J.~Matias, ``{Huge right-handed current effects in $B \rightarrow
  K*(K \pi)l^+l^-$ in supersymmetry},'' {\em JHEP}, vol.~0704, p.~058, 2007,
  hep-ph/0612166.

\bibitem{Mahmoudi:2007vz}
F.~Mahmoudi, ``{SuperIso: A Program for calculating the isospin asymmetry of $B
  \rightarrow K* \gamma$ in the MSSM},'' {\em Comput.Phys.Commun.}, vol.~178,
  pp.~745--754, 2008, 0710.2067.

\bibitem{Mahmoudi:2008tp}
F.~Mahmoudi, ``{SuperIso v2.3: A Program for calculating flavor physics
  observables in Supersymmetry},'' {\em Comput.Phys.Commun.}, vol.~180,
  pp.~1579--1613, 2009, 0808.3144.

\bibitem{Chatrchyan:2012rg}
S.~Chatrchyan {\em et~al.}, ``{Search for $B^0_s to \mu^+ \mu^-$ and $B^0 to
  \mu^+ \mu^-$ decays},'' {\em JHEP}, vol.~1204, p.~033, 2012, 1203.3976.

\bibitem{Aaij:2012ac}
R.~Aaij {\em et~al.}, ``{Strong constraints on the rare decays $B_s \to \mu^+
  \mu^-$ and $B^0 \to \mu^+ \mu^-$},'' {\em Phys.Rev.Lett.}, vol.~108,
  p.~231801, 2012, 1203.4493.

\bibitem{Hryczuk:2010zi}
A.~Hryczuk, R.~Iengo, and P.~Ullio, ``{Relic densities including Sommerfeld
  enhancements in the MSSM},'' {\em JHEP}, vol.~1103, p.~069, 2011, 1010.2172.

\bibitem{ArkaniHamed:2006mb}
N.~Arkani-Hamed, A.~Delgado, and G.~Giudice, ``{The Well-tempered
  neutralino},'' {\em Nucl.Phys.}, vol.~B741, pp.~108--130, 2006,
  hep-ph/0601041.

\bibitem{Ellis:2005mb}
J.~R. Ellis, K.~A. Olive, Y.~Santoso, and V.~C. Spanos, ``{Update on the direct
  detection of supersymmetric dark matter},'' {\em Phys.Rev.}, vol.~D71,
  p.~095007, 2005, hep-ph/0502001.

\bibitem{Drees:1992rr}
M.~Drees and M.~M. Nojiri, ``{New contributions to coherent neutralino -
  nucleus scattering},'' {\em Phys.Rev.}, vol.~D47, pp.~4226--4232, 1993,
  hep-ph/9210272.

\bibitem{Drees:1993bu}
M.~Drees and M.~Nojiri, ``{Neutralino - nucleon scattering revisited},'' {\em
  Phys.Rev.}, vol.~D48, pp.~3483--3501, 1993, hep-ph/9307208.

\bibitem{BT:1998}
J.~Binney and S.~Tremaine, ``{Galactic Dynamics: Second Edition},'' {\em
  Princeton University Press}, 2008.

\bibitem{Freudenreich:1997bx}
H.~Freudenreich, ``{Cobe's galactic bar and disk},'' {\em Astrophys.J.},
  vol.~492, pp.~495--510, 1998, astro-ph/9707340.

\bibitem{dame:267}
T.~M. Dame, ``The distribution of neutral gas in the milky way,'' {\em AIP
  Conference Proceedings}, vol.~278, no.~1, pp.~267--278, 1992.

\bibitem{Navarro:2003ew}
J.~F. Navarro, E.~Hayashi, C.~Power, A.~Jenkins, C.~S. Frenk, {\em et~al.},
  ``{The Inner structure of Lambda-CDM halos 3: Universality and asymptotic
  slopes},'' {\em Mon.Not.Roy.Astron.Soc.}, vol.~349, p.~1039, 2004,
  astro-ph/0311231.

\bibitem{Graham:2006af}
A.~W. Graham, D.~Merritt, B.~Moore, J.~Diemand, and B.~Terzic, ``{Empirical
  Models for Dark Matter Halos. III. The Kormendy relation and the
  $log(rho_e)-log(R_e)$ relation},'' {\em Astron.J.}, vol.~132, pp.~2711--2716,
  2006, astro-ph/0608614.

\bibitem{Bryan:1997dn}
G.~Bryan and M.~Norman, ``{Statistical properties of x-ray clusters: Analytic
  and numerical comparisons},'' {\em Astrophys.J.}, vol.~495, p.~80, 1998,
  astro-ph/9710107.

\bibitem{Komatsu:2010fb}
E.~Komatsu {\em et~al.}, ``{Seven-Year Wilkinson Microwave Anisotropy Probe
  (WMAP) Observations: Cosmological Interpretation},'' {\em
  Astrophys.J.Suppl.}, vol.~192, p.~18, 2011, 1001.4538.

\bibitem{Xue:2008se}
X.~Xue {\em et~al.}, ``{The Milky Way's Circular Velocity Curve to 60 kpc and
  an Estimate of the Dark Matter Halo Mass from Kinematics of ~2400 SDSS Blue
  Horizontal Branch Stars},'' {\em Astrophys.J.}, vol.~684, pp.~1143--1158,
  2008, 0801.1232.

\bibitem{Akrami:2010dn}
Y.~Akrami, C.~Savage, P.~Scott, J.~Conrad, and J.~Edsjo, ``{How well will
  ton-scale dark matter direct detection experiments constrain minimal
  supersymmetry?},'' {\em JCAP}, vol.~1104, p.~012, 2011, 1011.4318.

\bibitem{Bertone:2011nj}
G.~Bertone, D.~G. Cerdeno, M.~Fornasa, R.~Ruiz~de Austri, C.~Strege, {\em
  et~al.}, ``{Global fits of the cMSSM including the first LHC and XENON100
  data},'' {\em JCAP}, vol.~1201, p.~015, 2012, 1107.1715.

\bibitem{Cerdeno:2012ix}
D.~G. Cerdeno, M.~Fornasa, J.-H. Huh, and M.~Peiro, ``{Nuclear uncertainties in
  the spin-dependent structure functions for direct dark matter detection},''
  2012, 1208.6426.

\bibitem{Akrami:2010cz}
Y.~Akrami, C.~Savage, P.~Scott, J.~Conrad, and J.~Edsjo, ``{Statistical
  coverage for supersymmetric parameter estimation: a case study with direct
  detection of dark matter},'' {\em JCAP}, vol.~1107, p.~002, 2011, 1011.4297.

\bibitem{Trotta:2008bp}
R.~Trotta, F.~Feroz, M.~P. Hobson, L.~Roszkowski, and R.~Ruiz~de Austri, ``{The
  Impact of priors and observables on parameter inferences in the Constrained
  MSSM},'' {\em JHEP}, vol.~0812, p.~024, 2008, 0809.3792.

\bibitem{Bertone:2010rv}
G.~Bertone, D.~G. Cerdeno, M.~Fornasa, R.~R. de~Austri, and R.~Trotta,
  ``{Identification of Dark Matter particles with LHC and direct detection
  data},'' {\em Phys.Rev.}, vol.~D82, p.~055008, 2010, 1005.4280.

\bibitem{Gelmini:2006mr}
G.~B. Gelmini, P.~Gondolo, A.~Soldatenko, and C.~E. Yaguna, ``{Direct detection
  of neutralino dark mattter in non- standard cosmologies},'' {\em Phys. Rev.},
  vol.~D76, p.~015010, 2007, hep-ph/0610379.

\bibitem{Acharya:2009zt}
B.~S. Acharya, G.~Kane, S.~Watson, and P.~Kumar, ``{A Non-thermal WIMP
  Miracle},'' {\em Phys. Rev.}, vol.~D80, p.~083529, 2009, 0908.2430.

\bibitem{Arcadi:2011ev}
G.~Arcadi and P.~Ullio, ``{Accurate estimate of the relic density and the
  kinetic decoupling in non-thermal dark matter models},'' {\em Phys.Rev.},
  vol.~D84, p.~043520, 2011, 1104.3591.

\bibitem{Catena:2004ba}
R.~Catena, N.~Fornengo, A.~Masiero, M.~Pietroni, and F.~Rosati, ``{Dark matter
  relic abundance and scalar - tensor dark energy},'' {\em Phys.Rev.},
  vol.~D70, p.~063519, 2004, astro-ph/0403614.

\bibitem{Catena:2007ix}
R.~Catena, N.~Fornengo, A.~Masiero, M.~Pietroni, and M.~Schelke, ``{Enlarging
  mSUGRA parameter space by decreasing pre-BBN Hubble rate in Scalar-Tensor
  Cosmologies},'' {\em JHEP}, vol.~0810, p.~003, 2008, 0712.3173.

\bibitem{Aad:2009wy}
G.~Aad {\em et~al.}, ``{Expected Performance of the ATLAS Experiment -
  Detector, Trigger and Physics},'' 2009, 0901.0512.

\bibitem{Alwall:2011uj}
J.~Alwall, M.~Herquet, F.~Maltoni, O.~Mattelaer, and T.~Stelzer, ``{MadGraph 5
  : Going Beyond},'' {\em JHEP}, vol.~1106, p.~128, 2011, 1106.0522.

\bibitem{Beenakker:1996ed}
W.~Beenakker, R.~Hopker, and M.~Spira, ``{PROSPINO: A Program for the
  production of supersymmetric particles in next-to-leading order QCD},'' 1996,
  hep-ph/9611232.

\bibitem{Aad:3l}
G.~Aad {\em et~al.}, ``{Search for direct production of charginos and
  neutralinos in events with three leptons and missing transverse momentum in
  sqrt(s) = 7 TeV pp collisions with the ATLAS detector},'' 2012, 1208.3144.

\bibitem{Aad:2l}
G.~Aad {\em et~al.}, ``{Search for direct slepton and gaugino production in
  final states with two leptons and missing transverse momentum with the ATLAS
  detector in pp collisions at sqrt(s) = 7 TeV},'' 2012, 1208.2884.

\bibitem{ATLAS-CONF-2012-084}
``Search for dark matter candidates and large extra dimensions in events with a
  jet and missing transverse momentum with the atlas detector,'' Tech. Rep.
  ATLAS-CONF-2012-084, CERN, Geneva, Jul 2012.

\bibitem{:2012rz}
G.~Aad {\em et~al.}, ``{Search for squarks and gluinos with the ATLAS detector
  in final states with jets and missing transverse momentum using 4.7 $fb^-1$
  of sqrt(s) = 7 TeV proton-proton collision data},'' 2012, 1208.0949.

\bibitem{ATLAS-CONF-2012-109}
``Search for squarks and gluinos with the atlas detector using final states
  with jets and missing transverse momentum and 5.8 fb$^{-1}$ of $\sqrt{s}$=8
  tev proton-proton collision data,'' Tech. Rep. ATLAS-CONF-2012-109, CERN,
  Geneva, Aug 2012.

\bibitem{Baer:2012ts}
H.~Baer, V.~Barger, A.~Lessa, W.~Sreethawong, and X.~Tata, ``{Wh plus
  missing-$E_T $signature from gaugino pair production at the LHC},'' {\em
  Phys.Rev.}, vol.~D85, p.~055022, 2012, 1201.2949.

\bibitem{Conley:2011nn}
J.~A. Conley, J.~S. Gainer, J.~L. Hewett, M.~P. Le, and T.~G. Rizzo,
  ``{Supersymmetry Without Prejudice at the 7 TeV LHC},'' {\em Physical Review
  D}, 2011, 1103.1697.

\bibitem{Dreiner:2012gx}
H.~K. Dreiner, M.~Kramer, and J.~Tattersall, ``{How low can SUSY go? Matching,
  monojets and compressed spectra},'' {\em Europhys.Lett.}, vol.~99, p.~61001,
  2012, 1207.1613.

\bibitem{Belanger:2012mk}
G.~Belanger, M.~Heikinheimo, and V.~Sanz, ``{Model-Independent Bounds on
  Squarks from Monophoton Searches},'' {\em JHEP}, vol.~1208, p.~151, 2012,
  1205.1463.

\bibitem{Aad:2011cw}
G.~Aad {\em et~al.}, ``{Search for scalar bottom pair production with the ATLAS
  detector in pp Collisions at sqrt{s} = 7 TeV},'' {\em Phys.Rev.Lett.},
  vol.~108, p.~181802, 2012, 1112.3832.

\bibitem{ATLAS-CONF-2012-106}
``Search for scalar bottom pair production in Þnal states with missing
  transverse momentum and two b-jets in pp collisions at Ãs=7 tev with the
  atlas detector,'' Tech. Rep. ATLAS-CONF-2012-106, CERN, Geneva, Aug 2012.

\bibitem{Collaboration:2012ar}
G.~Aad {\em et~al.}, ``{Search for direct top squark pair production in final
  states with one isolated lepton, jets, and missing transverse momentum in
  sqrt(s) = 7 TeV pp collisions using 4.7 fb-1 of ATLAS data},'' 2012,
  1208.2590.

\bibitem{Collaboration:2012si}
G.~Aad {\em et~al.}, ``{Search for a supersymmetric partner to the top quark in
  final states with jets and missing transverse momentum at sqrt(s) = 7 TeV
  with the ATLAS detector},'' 2012, 1208.1447.

\bibitem{:2012yr}
G.~Aad {\em et~al.}, ``{Search for light top squark pair production in final
  states with leptons and b-jets with the ATLAS detector in sqrt(s) = 7 TeV
  proton-proton collisions},'' 2012, 1209.2102.

\bibitem{Choudhury:2012tc}
A.~Choudhury and A.~Datta, ``{Many faces of low mass neutralino dark matter in
  the unconstrained MSSM, LHC data and new signals},'' {\em JHEP}, vol.~1206,
  p.~006, 2012, 1203.4106.

\bibitem{Adriani:2010rc}
O.~Adriani {\em et~al.}, ``{PAMELA results on the cosmic-ray antiproton flux
  from 60 MeV to 180 GeV in kinetic energy},'' {\em Phys.Rev.Lett.}, vol.~105,
  p.~121101, 2010, 1007.0821.

\bibitem{Bringmann:2012vr}
T.~Bringmann, X.~Huang, A.~Ibarra, S.~Vogl, and C.~Weniger, ``{Fermi LAT Search
  for Internal Bremsstrahlung Signatures from Dark Matter Annihilation},'' {\em
  JCAP}, vol.~1207, p.~054, 2012, 1203.1312.

\bibitem{Weniger:2012tx}
C.~Weniger, ``{A Tentative Gamma-Ray Line from Dark Matter Annihilation at the
  Fermi Large Area Telescope},'' {\em JCAP}, vol.~1208, p.~007, 2012,
  1204.2797.

\bibitem{Bringmann:2012ez}
T.~Bringmann and C.~Weniger, ``{Gamma Ray Signals from Dark Matter: Concepts,
  Status and Prospects},'' {\em Phys.Dark Univ.}, vol.~1, pp.~194--217, 2012,
  1208.5481.

\bibitem{Buchmuller:2012rc}
W.~Buchmuller and M.~Garny, ``{Decaying vs Annihilating Dark Matter in Light of
  a Tentative Gamma-Ray Line},'' {\em JCAP}, vol.~1208, p.~035, 2012,
  1206.7056.

\bibitem{Aad:2012ky}
G.~Aad {\em et~al.}, ``{Search for dark matter candidates and large extra
  dimensions in events with a jet and missing transverse momentum with the
  ATLAS detector},'' 2012, 1210.4491.

\bibitem{Choudhury:2012kn}
A.~Choudhury and A.~Datta, ``{New limits on top squark NLSP from LHC 4.7
  $fb^{-1}$ data},'' 2012, 1207.1846.

\end{thebibliography}
\bibliographystyle{hieeetr}

\end{document}